\documentclass[10pt]{article}
\usepackage[dvips]{graphicx}
\usepackage{amssymb}

\usepackage{cite}

\title{Dynamic structure factors of the spin-$1/2$ $XX$ chain
       \protect\\
       with Dzyaloshinskii-Moriya interaction}
\author{Oleg Derzhko$^{1,2}$
        and
        Taras Verkholyak$^1$\\
\small{$^1$Institute for Condensed Matter Physics,
       National Academy of Sciences of Ukraine,}\\
\small{1 Svientsitskii Street, L'viv-11, 79011, Ukraine}\\
\small{$^2$The Abdus Salam International Centre for Theoretical Physics,}\\
\small{Strada Costiera 11, Trieste, 34100, Italy}}

\date{\today}

\begin{document}

\renewcommand\baselinestretch {1.25}
\large\normalsize

\maketitle

\begin{abstract}
We consider the spin-$1/2$ isotropic $XY$ chain in a ($z$) transverse magnetic field
with the Dzya\-lo\-shin\-skii-Moriya interaction
directed along the $z$-axis in spin space
and examine the effects of the latter interaction
on the $zz$, $xx$ ($yy$) and $xy$ ($yx$) dynamic structure factors.
The Dzyaloshinskii-Moriya interaction does not manifest itself
in the $zz$ dynamic quantities.
In contrast,
the $xx$ ($yy$) and $xy$ ($yx$) dynamic structure factors
show dramatical changes owing to the Dzyaloshinskii-Moriya interaction.
Implications of our results for electron spin resonance experiments
are briefly discussed.
\end{abstract}

\vspace{2mm}

\noindent
{\bf PACS number(s):}
75.10.-b

\vspace{2mm}

\noindent
{\bf Keywords:}
spin-$1/2$ $XY$ chain,
Dzyaloshinskii-Moriya interaction,
dynamic structure factors,
ESR

\vspace{5mm}

\renewcommand\baselinestretch {1.35}
\large\normalsize

\section{Introductory remarks}
\label{s1}

Magnetism in low-dimensional compounds
which are modeled by quantum spin systems
has been a subject of intense study in recent years.
Among the experimental techniques
which are used to probe various characteristics of magnetic systems
the dynamic experiments are very important.
Due to their high sensitivity and resolution
they are among the basic tools to determine the detailed magnetic interactions.
However, the correct interpretation of the experimental data
requires a corresponding theoretical background.
Usually the theories
based on mean-field approximation or classical spin picture
are not sufficient to explain dynamic phenomena
in low-dimensional quantum spin systems
in which quantum fluctuations may be very strong
(see e.g. \cite{01,02,03,04,05}).

In the present work we examine rigorously the dynamic properties
of a quantum spin chain with the Dzya\-lo\-shin\-skii-Moriya interaction.
The latter interaction is often present in low-dimensional magnetic insulators.
Although its value is usually small
it may cause noticeable changes
in different observable characteristics of magnetic systems
(see e.g. \cite{06}).
In our study we restrict ourselves
to the spin-$1/2$ isotropic $XY$
(i.e. $XX$ or $XX0$)
chain in a ($z$) transverse magnetic field
with the Dzyaloshinskii-Moriya interaction
directed along $z$-axis in spin space
\cite{07,08}.
That model is equivalent to a chain of free spinless fermions.
In fermionic language,
the Ising component of exchange coupling
(or other components of the Dzyaloshinskii-Moriya interaction
as well as the external magnetic field
directed along other axes in spin space)
corresponds to interaction between spinless fermions
and makes the problem much more complicated.
Although the spin-$1/2$ $XY$ chain
can be mapped onto a free spinless fermion chain,
the analysis of the dynamic characteristics of the spin model
is not trivial at all.
The $zz$ spin correlation function
corresponds to the density-density correlation function for spinless fermions
and its calculation is relatively easy \cite{09,10}.
In contrast,
the other ($xx$, $yy$, $xy$, $yx$) spin correlation functions
are more complicated quantities
due to the nonlocal character of the Jordan-Wigner relation
between the spin raising/lowering operators
and the creation/annihilation fermion operators.
The explicit analytical results for them
are available in the high-temperature limit \cite{11}
and in the ground state for strong fields \cite{12}.
From Refs. \cite{13,14} we know also the asymptotic behavior
of some time-dependent spin correlation functions
in particular limiting cases.
On the other hand,
the time-dependent spin correlation functions
(and the related dynamic structure factors and dynamic susceptibilities)
can be computed numerically \cite{15a,15b,15c}.

In the present paper,
we employ the Jordan-Wigner fermionization
supplemented by further analytical and numerical calculations
to perform a comprehensive analysis
of the effect of the Dzyaloshinskii-Moriya interaction
on the dynamic structure factors
of the spin-$1/2$ $XX$ chain in a transverse field.
The previous studies of the effect of the Dzyaloshinskii-Moriya interaction
were restricted to transverse ($zz$) dynamics \cite{16,17}
and to the dynamics of correlations between $x$ and $y$ spin components
at infinite temperature \cite{18}.
In particular,
the transverse dynamic susceptibility $\chi_{zz}(\kappa,\omega)$
was found at $\kappa=0$ \cite{16} and $\kappa\ne 0$ \cite{17}.
We also notice
that a relation between the $XY$ chains without and with the Dzyaloshinskii-Moriya interaction
on the grounds of symmetry was discussed in \cite{19}.
However,
the transverse dynamics was not discussed
in the context of the two-fermion excitation continuum \cite{10}.
Moreover,
to our best knowledge,
the effect of the Dzyaloshinskii-Moriya interaction
on the $xx$, $yy$, $xy$, $yx$ dynamic structure factors
away from the infinite temperature limit
has never been studied before.
We must note here
that the dynamic properties of a more general model,
the spin-$1/2$ antiferromagnetic $XXX$ Heisenberg chain,
with the Dzyaloshinskii-Moriya interaction
were explored in Refs. \cite{20,21}.
Thus,
in Ref. \cite{20},
using symmetry arguments it is shown
that although the Dzyaloshinskii-Moriya interaction
may leave the spectrum of the problem unchanged,
it can essentially influence the spin correlations / dynamic susceptibilities.
In Ref. \cite{21},
the spin correlation functions / dynamic susceptibilities
of the chain with the Dzyaloshinskii-Moriya interaction
are expressed in terms of such quantities of a $XXZ$ chain.
However,
the results for the latter chain are restricted to the limit
of wavevectors close to $\pi$, low energies and low temperatures.

The outline of this paper is as follows.
In the next section
(Sec. \ref{s2})
we introduce the model,
define the quantities of interest
and make some remarks on symmetry.
Then, in the next two sections
(Secs. \ref{s3} and \ref{s4})
we analyze the $zz$ dynamic structure factor
and
the $xx$ ($yy$), $xy$ ($yx$) dynamic structure factors,
respectively.
In Sec. \ref{s5}
we discuss the effect of the Dzyaloshinskii-Moriya interaction
on the electron spin resonance absorption spectrum.
In the last section
(Sec. \ref{s6})
we briefly summarize our findings.
Some preliminary results of the present paper
are reported in the conference papers \cite{22}.

\section{The model and its symmetries}
\label{s2}

The model to be studied is a one-dimensional lattice of $N\to\infty$ spins $1/2$.
The Hamiltonian of the model is given by
\begin{eqnarray}
H=\sum_{n=1}^NJ\left(s_n^xs_{n+1}^x+s_n^ys_{n+1}^y\right)
+\sum_{n=1}^ND\left(s_n^xs_{n+1}^y-s_n^ys_{n+1}^x\right)
-\sum_{n=1}^N h s_n^z,
\label{01}
\end{eqnarray}
where $s^{\alpha}$ are the halves of the Pauli matrices,
$J$ is the $XX$ exchange interaction,
$D$ is the Dzyaloshinskii-Moriya interaction,
and
$h$ is the transverse magnetic field.
We imply in (\ref{01})
both
periodic boundary conditions
(in analytical computations)
and
open boundary conditions
(in numerical computations)
bearing in mind that the results
which pertain to the thermodynamic limit $N\to\infty$
are insensitive to the boundary conditions imposed.
We are interested in the dynamic structure factors
of the model (\ref{01})
which can be expressed through the time-dependent spin correlation functions as follows
\begin{eqnarray}
S_{\alpha\beta}(\kappa,\omega)
=
\sum_{m=1}^N\exp\left(-{\rm{i}}\kappa m\right)
\int_{-\infty}^{\infty}{\mbox{d}}t\exp\left({\rm{i}}\omega t\right)
\left(\langle s_n^{\alpha}(t)s_{n+m}^{\beta}\rangle
-\langle s_n^{\alpha}\rangle\langle s_{n+m}^{\beta}\rangle\right).
\label{02}
\end{eqnarray}
Here
$\langle(\ldots)\rangle
={\rm{Tr}}\left(\exp(-\beta H)(\ldots)\right)
/{\rm{Tr}}\exp\left(-\beta H\right)$
denotes the canonical thermodynamic average
and $s_n^{\alpha}(t)=\exp({\rm{i}}Ht)s_n^{\alpha}\exp(-{\rm{i}}Ht)$
is the Heisenberg representation of the operator $s_n^{\alpha}$.
Note that
the diagonal components of $S_{\alpha\beta}(\kappa,\omega)$,
$S_{\alpha\alpha}(\kappa,\omega)$,
are real and nonnegative
but the off-diagonal components of $S_{\alpha\beta}(\kappa,\omega)$,
$\alpha\ne\beta$
should not necessarily be real.

First of all we note
that the spin model (\ref{01}) possesses a number of symmetries
which permit to reduce the range of parameters
and simplify studies of the properties of the model.
After performing a $\pi/2$-rotation of all spins about the $z$-axis we conclude
that
$S_{xx}(\kappa,\omega)=S_{yy}(\kappa,\omega)$
and
$S_{xy}(\kappa,\omega)=-S_{yx}(\kappa,\omega)$
and therefore,
the only dynamic structure factors
which remain to examine are
$S_{zz}(\kappa,\omega)$,
$S_{xx}(\kappa,\omega)$
and
$S_{xy}(\kappa,\omega)$.
We also note,
that making use of the transformation
$\tilde{s}_n^x=s_n^x$,
$\tilde{s}_n^y=-s_n^y$,
$\tilde{s}_n^z=-s_n^z$
we arrive at (\ref{01}) with the parameters $J$, $-D$, $-h$
and as a result,
$S_{\alpha\alpha}(\kappa,\omega)\vert_{D,h}
=S_{\alpha\alpha}(\kappa,\omega)\vert_{-D,-h}$,
$\alpha=x,y,z$
but
$S_{xy}(\kappa,\omega)\vert_{D,h}
=-S_{xy}(\kappa,\omega)\vert_{-D,-h}$.
A similar transformation,
$\tilde{s}_n^x=(-1)^ns_n^x$,
$\tilde{s}_n^y=(-1)^ns_n^y$,
$\tilde{s}_n^z=s_n^z$,
yields (\ref{01}) with the parameters $-J$, $-D$, $h$;
this symmetry implies that
$S_{\alpha\beta}(\kappa,\omega)\vert_{-J,-D}
=S_{\alpha\beta}(\kappa\pm\pi,\omega)\vert_{J,D}$,
$\alpha,\beta=x,y$,
$S_{zz}(\kappa,\omega)\vert_{-J,-D}
=S_{zz}(\kappa,\omega)\vert_{J,D}$.
Finally,
the renumbering of sites
$j\to N-j+1$, $j=1,\ldots,N$
gives again (\ref{01}) with the parameters $J$, $-D$, $h$
and hence
$S_{\alpha\beta}(\kappa,\omega)\vert_{D}
=S_{\alpha\beta}(-\kappa,\omega)\vert_{-D}$,
$\alpha,\beta=x,y,z$.

In our study an important role is played by
a gauge transformation
that eliminates the Dzya\-lo\-shin\-skii-Moriya interaction
from the Hamiltonian (\ref{01})
at the expense of renormalized $XX$ exchange interaction
(see e.g. \cite{06,19,20,21,23,24}).
Explicitly,
this spin coordinate transformation reads
\begin{eqnarray}
\tilde{s}_n^x=s_n^x\cos\phi_n+s_n^y\sin\phi_n,
\;\;\;
\tilde{s}_n^y=-s_n^x\sin\phi_n+s_n^y\cos\phi_n,
\;\;\;
\tilde{s}_n^z=s_n^z,
\;\;\;
\phi_n=(n-1)\varphi,
\;\;\;
\tan\varphi=\frac{D}{J}.
\label{03}
\end{eqnarray}
With the aid of (\ref{03}) we find
that the Hamiltonian (\ref{01}) becomes
\begin{eqnarray}
H=\sum_{n=1}^N\tilde{J}
\left(\tilde{s}_n^x\tilde{s}_{n+1}^x+\tilde{s}_n^y\tilde{s}_{n+1}^y\right)
-\sum_{n=1}^N h \tilde{s}_n^z,
\;\;\;
\tilde{J}={\rm{sgn}}(J)\sqrt{J^2+D^2}.
\label{04}
\end{eqnarray}
Eq. (\ref{04}) corresponds to the model without the Dzyaloshinskii-Moriya interaction,
however,
with the renormalized $XX$ exchange interaction $\tilde{J}$.
In what follows 
we use the large body of existing knowledge about the model (\ref{04})
to obtain the properties of the model with the Dzyaloshinskii-Moriya interaction (\ref{01})
exploiting the transformation (\ref{03}).
In principle,
no new calculations
(in comparison with those reported in \cite{15c})
are needed.
Moreover,
in Sec. \ref{s4} we make use of the results for $S_{xy}(\kappa,\omega)$
obtained in Ref. \cite{15c},
the physical meaning of which were nebulous and not obvious earlier.
To close this section,
we note that
the Dzyaloshinskii-Moriya interaction cannot be detected
from measurements of the thermodynamic quantities
since these quantities cannot distinguish between the models (\ref{01}) and (\ref{04})
related through the unitary transformation (\ref{03}).
However,
as we shall see below,
the Dzyaloshinskii-Moriya interaction can be determined from some of the dynamic quantities.

\section{$zz$ dynamic structure factor}
\label{s3}

We start with the transverse dynamic structure factor.
Since the spin rotations (\ref{03}) do not affect the $z$ spin component
we immediately obtain the following expression
for the transverse dynamic structure factor of the spin chain (\ref{01})
\begin{eqnarray}
S_{zz}(\kappa,\omega)
=
\int_{-\pi}^{\pi}{\rm{d}}\kappa_1
n_{\kappa_1}\left(1-n_{\kappa_1+\kappa}\right)
\delta\left(\omega+\Lambda_{\kappa_1}-\Lambda_{\kappa_1+\kappa}\right),
\nonumber\\
\Lambda_{\kappa}=-h+\tilde{J}\cos\kappa,
\;\;\;
n_{\kappa}=\frac{1}{1+\exp\left(\beta\Lambda_{\kappa}\right)}
\label{05}
\end{eqnarray}
(see \cite{10} and references therein).

As can be seen from Eq. (\ref{05}),
the transverse dynamics
is conditioned by a continuum of two-fermion (particle-hole) excitations.
The properties of the two-fermion excitation continuum
were studied in detail in Refs. \cite{25,10}.
We need these results in what follows
and therefore present them here for easy reference.
For later convenience
we introduce the following characteristic lines in the $\kappa$--$\omega$ plane
\begin{eqnarray}
\frac{\omega^{(1)}(\kappa)}{\sqrt{J^2+D^2}}
=2\left\vert
\sin\frac{\kappa}{2}
\sin\left(\frac{\vert\kappa\vert}{2}-\alpha\right)
\right\vert,
\label{06}
\end{eqnarray}
\begin{eqnarray}
\frac{\omega^{(2)}(\kappa)}{\sqrt{J^2+D^2}}
=2\left\vert
\sin\frac{\kappa}{2}
\sin\left(\frac{\vert\kappa\vert}{2}+\alpha\right)
\right\vert,
\label{07}
\end{eqnarray}
\begin{eqnarray}
\frac{\omega^{(3)}(\kappa)}{\sqrt{J^2+D^2}}
=2\left\vert
\sin\frac{\kappa}{2}
\right\vert,
\label{08}
\end{eqnarray}
where the parameter
$\alpha=\arccos\left(\vert h\vert/\sqrt{J^2+D^2}\right)$
varies from $\pi/2$ when $h=0$
to $0$ when $\vert h\vert=\sqrt{J^2+D^2}$.
(We notice that the authors of Ref. \cite{25} used another parameter
$\sigma$,
$\pi\sigma=\pi/2-\alpha$.)

The ground-state transverse dynamic structure factor does not vanish
until $\vert h\vert<\sqrt{J^2+D^2}$
and may have nonzero values
only within a restricted region of the $\kappa$--$\omega$ plane
($\vert\kappa\vert\le\pi$, $\omega\ge 0$)
with the lower boundary
$\omega_l(\kappa)=\omega^{(1)}(\kappa)$
and the upper boundary
$\omega_u(\kappa)=\omega^{(2)}(\kappa)$
if $0\le\vert\kappa\vert\le\pi-2\alpha$
or
$\omega_u(\kappa)=\omega^{(3)}(\kappa)$
if $\pi-2\alpha\le\vert\kappa\vert\le\pi$.
As it follows from (\ref{06}),
the soft modes occur at $\vert\kappa_0\vert=0,\;2\alpha$.
Moreover,
the transverse dynamic structure factor
exhibits a finite jump
along the middle boundary
$\omega_m(\kappa)=\omega^{(2)}(\kappa)$,
$\pi-2\alpha\le\vert\kappa\vert\le\pi$.
Finally,
$S_{zz}(\kappa,\omega)$ diverges along the curve
$\omega_s(\kappa)=\omega^{(3)}(\kappa)$.
This is the van Hove density-of-states effect in one dimension.
With growing temperature
the lower boundary becomes smeared out and finally disappears.
The upper boundary at nonzero temperatures is given by $\omega^{(3)}(\kappa)$ (\ref{08})
and $S_{zz}(\kappa,\omega)$
exhibits the one-dimensional van Hove singularity along this boundary.
The transverse dynamic structure factor becomes field-independent
in the high-temperature limit.

Clearly,
the Dzyaloshinskii-Moriya interaction
does not manifest itself in the transverse dynamics.
Again we cannot distinguish
the quantities probing transverse spin dynamics
for the spin-$1/2$ $XX$ chain with the Dzyaloshinskii-Moriya interaction
and
for the spin-$1/2$ $XX$ chain without the Dzyaloshinskii-Moriya interaction
but with renormalized $XX$ exchange interaction.
In the next section
we show what particular dynamic characteristics can unambiguously indicate
the presence of the Dzyaloshinskii-Moriya interaction.

\section{$xx$ and $xy$ dynamic structure factors}
\label{s4}

We turn to the remaining dynamic structure factors.
Exploiting the transformation (\ref{03})
one finds the relations
between
the $xx$ and $xy$ dynamic structure factors (\ref{02}) of the model (\ref{01})
(the l.h.s. of Eqs. (\ref{09}), (\ref{10}))
and
the $xx$ and $xy$ dynamic structure factors (\ref{02}) of the model (\ref{04})
(the r.h.s. of Eqs. (\ref{09}), (\ref{10}))
\begin{eqnarray}
S_{xx}(\kappa,\omega)
=\frac{1}{2}
\left(
S_{xx}(\kappa-\varphi,\omega)\vert_{\tilde{J}}
+S_{xx}(\kappa+\varphi,\omega)\vert_{\tilde{J}}
+{\rm{i}}
\left(
S_{xy}(\kappa-\varphi,\omega)\vert_{\tilde{J}}
-S_{xy}(\kappa+\varphi,\omega)\vert_{\tilde{J}}
\right)
\right),
\label{09}
\end{eqnarray}
\begin{eqnarray}
S_{xy}(\kappa,\omega)
=\frac{1}{2}
\left(
S_{xy}(\kappa-\varphi,\omega)\vert_{\tilde{J}}
+S_{xy}(\kappa+\varphi,\omega)\vert_{\tilde{J}}
-{\rm{i}}
\left(
S_{xx}(\kappa-\varphi,\omega)\vert_{\tilde{J}}
-S_{xx}(\kappa+\varphi,\omega)\vert_{\tilde{J}}
\right)
\right).
\label{10}
\end{eqnarray}

Now we use Eqs. (\ref{09}) and (\ref{10})
and the long known results for dynamics of the model (\ref{04})
to explore the effect of the Dzyaloshinskii-Moriya interaction
on $xx$ and $xy$ dynamics.
We begin with the case of infinite temperature $\beta=0$ \cite{11}.
In this limit,
the $xx$ and $xy$ time-dependent spin correlations at different sites vanish
and only the autocorrelations survive,
$\langle s_n^x(t)s_{n+m}^x\rangle\vert_{\tilde{J}}
=(1/4)\cos(ht)\exp\left(-\tilde{J}^2t^2/4\right)\delta_{m,0}$,
$\langle s_n^x(t)s_{n+m}^y\rangle\vert_{\tilde{J}}
=(1/4)\sin(ht)\exp\left(-\tilde{J}^2t^2/4\right)\delta_{m,0}$
(in that case
(i.e. when only autocorrelations survive)
according to (\ref{03})
we have simply
$\langle s_n^x(t)s_{n+m}^x\rangle\vert_{J,D}
=\langle s_n^x(t)s_{n+m}^x\rangle\vert_{\tilde{J}}$
and
$\langle s_n^x(t)s_{n+m}^y\rangle\vert_{J,D}
=\langle s_n^x(t)s_{n+m}^y\rangle\vert_{\tilde{J}}$;
this result for the model (\ref{01}),
which follows by symmetry
from the results for the model (\ref{04}),
was also derived earlier by direct calculation \cite{18}
(see Eq. (4.8) of that paper)).
Since the $xx$ and $xy$ dynamic structure factors do not depend on $\kappa$
the Dzyaloshinskii-Moriya interaction 
in accordance with (\ref{09}), (\ref{10})
changes only the energy scale
for these dynamic structure factors in the high-temperature limit $\beta=0$.
Explicitly,
the closed-form expressions for these quantities read
\begin{eqnarray}
S_{xx}(\kappa,\omega)
=
\frac{\sqrt{\pi}}{4\tilde{J}}
\left(
\exp\left(-\frac{\left(\omega+h\right)^2}{{\tilde{J}}^2}\right)
+
\exp\left(-\frac{\left(\omega-h\right)^2}{{\tilde{J}}^2}\right)
\right)
\label{11}
\end{eqnarray}
and
\begin{eqnarray}
{\rm{i}}S_{xy}(\kappa,\omega)
=
\frac{\sqrt{\pi}}{4\tilde{J}}
\left(
\exp\left(-\frac{\left(\omega+h\right)^2}{{\tilde{J}}^2}\right)
-
\exp\left(-\frac{\left(\omega-h\right)^2}{{\tilde{J}}^2}\right)
\right).
\label{12}
\end{eqnarray}
Thus,
in the high-temperature limit $\beta=0$
the $xx$ and $xy$ dynamic structure factors are $\kappa$-independent
and display a single Gaussian ridge at $\omega=\vert h\vert$.

\begin{figure}
\begin{center}
\includegraphics[clip=on,width=60mm,height=35mm,angle=0]{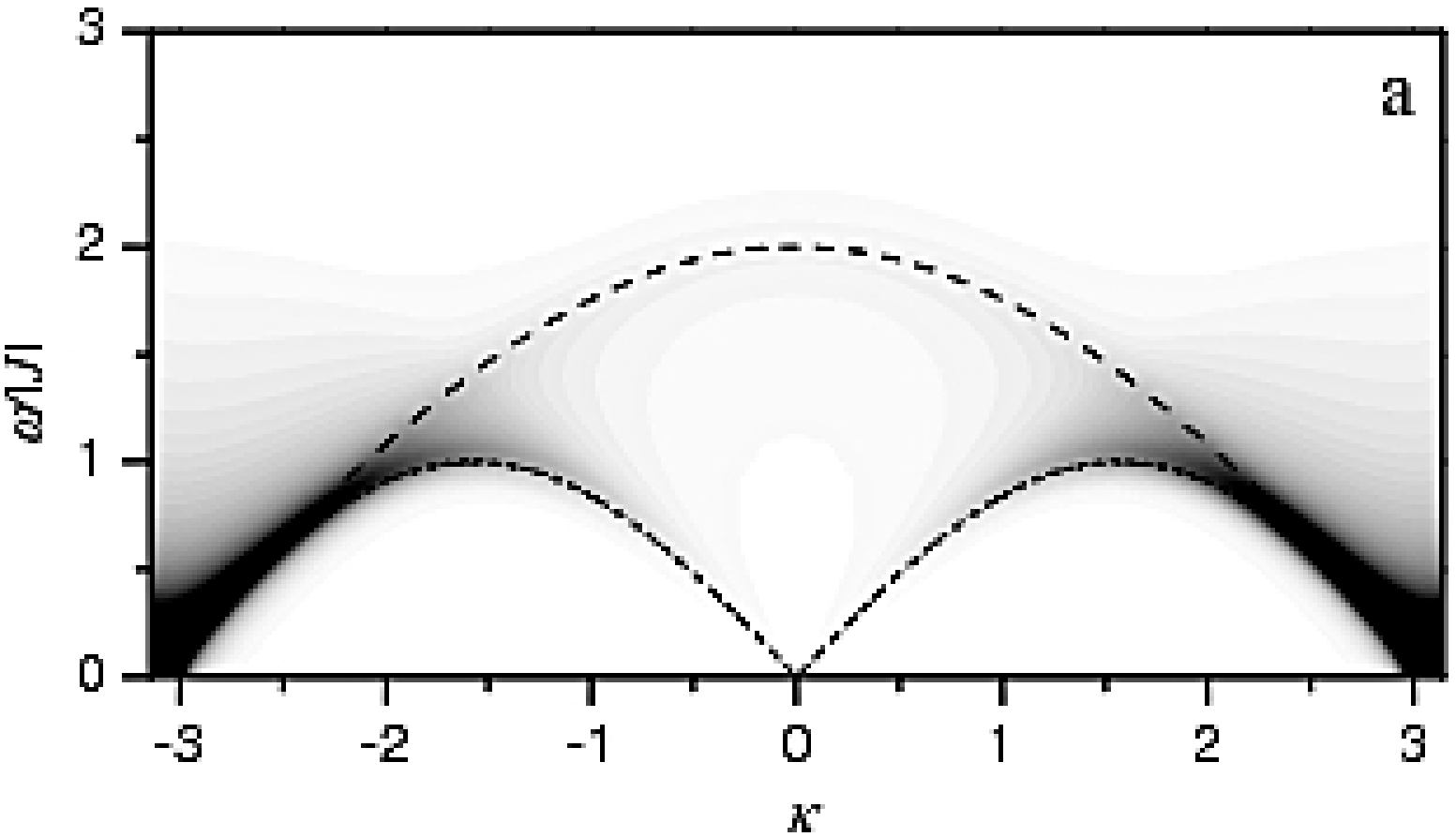}
\includegraphics[clip=on,width=60mm,height=35mm,angle=0]{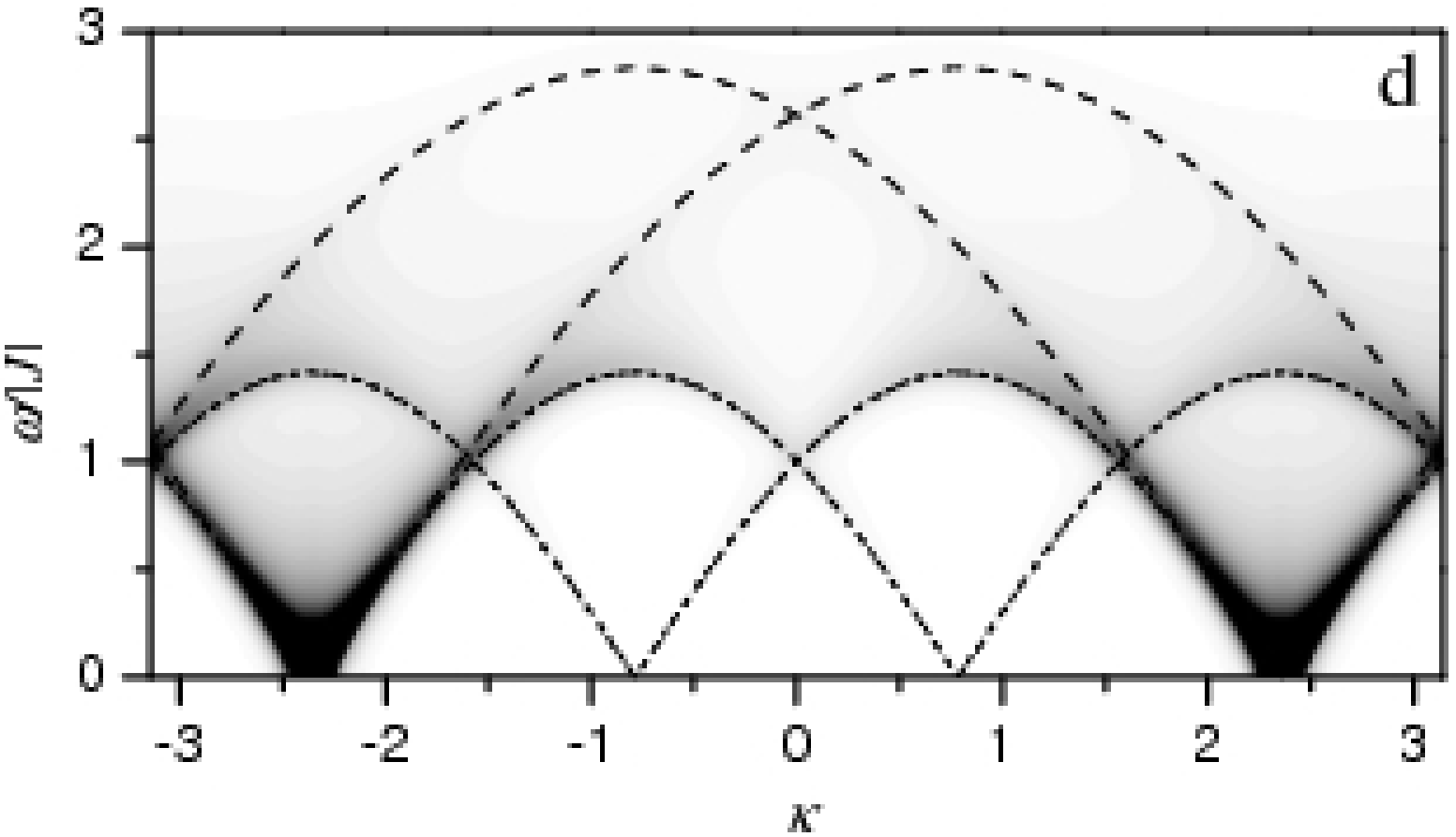}
\linebreak
\includegraphics[clip=on,width=60mm,height=35mm,angle=0]{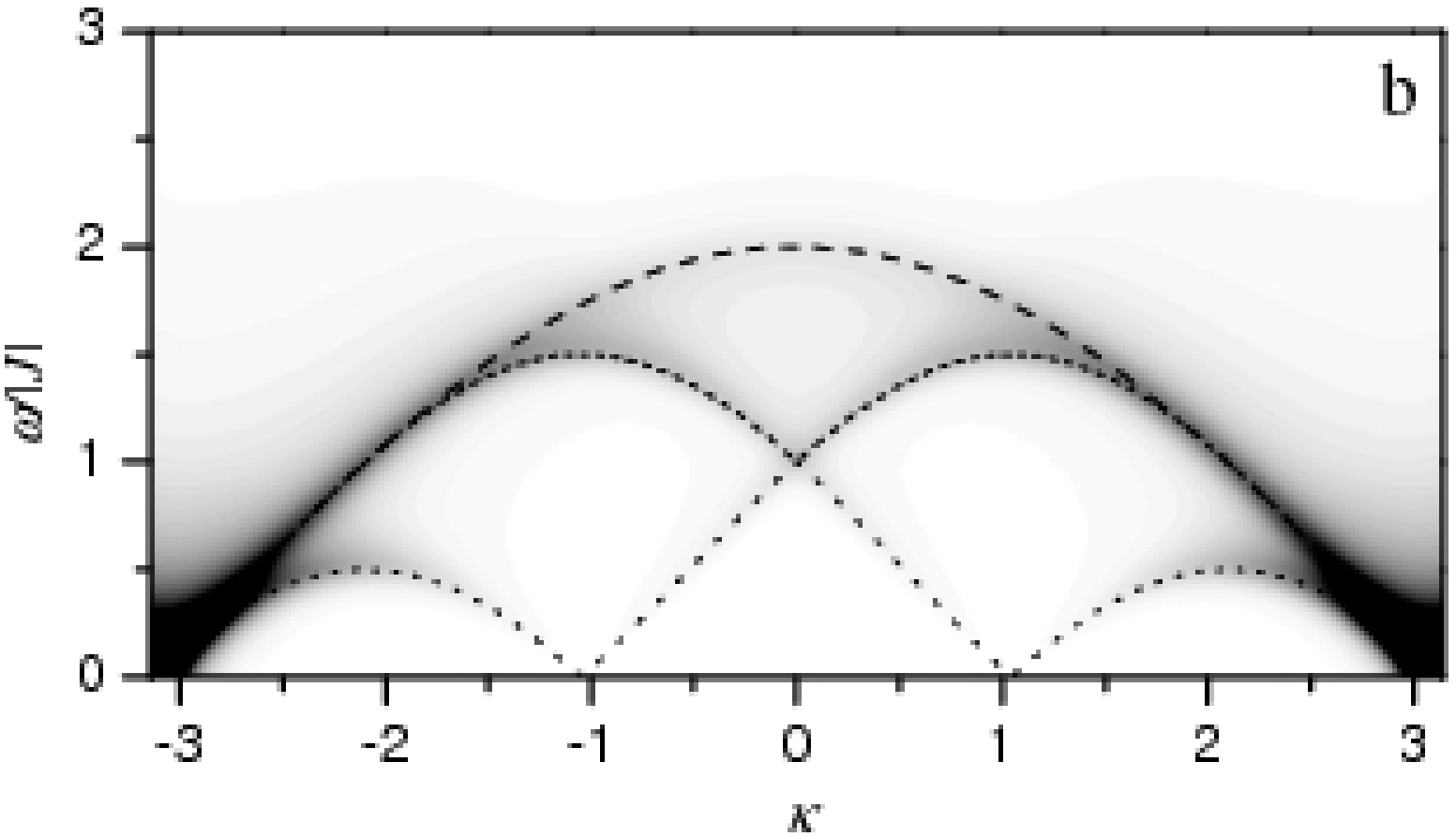}
\includegraphics[clip=on,width=60mm,height=35mm,angle=0]{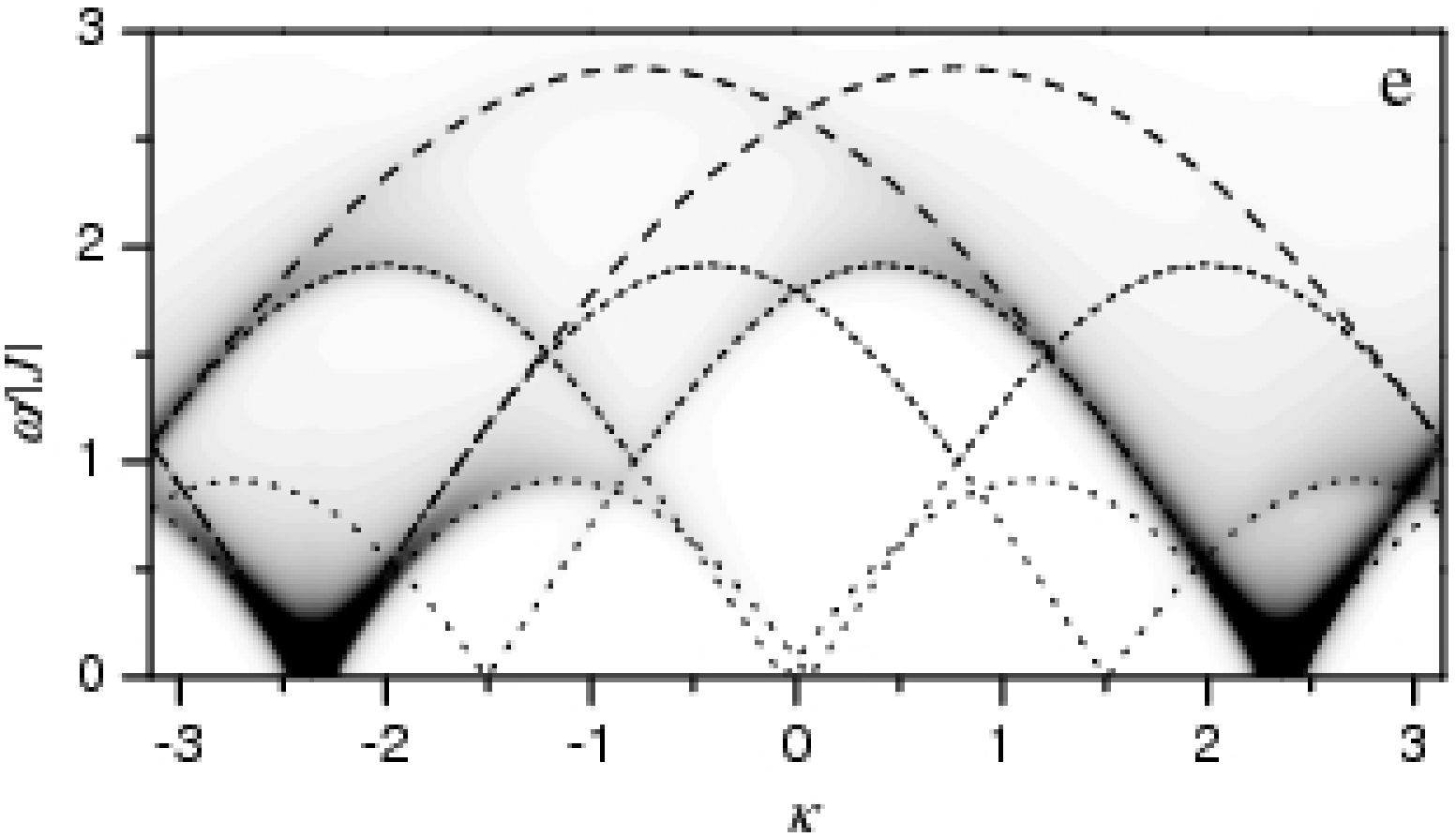}
\linebreak
\includegraphics[clip=on,width=60mm,height=35mm,angle=0]{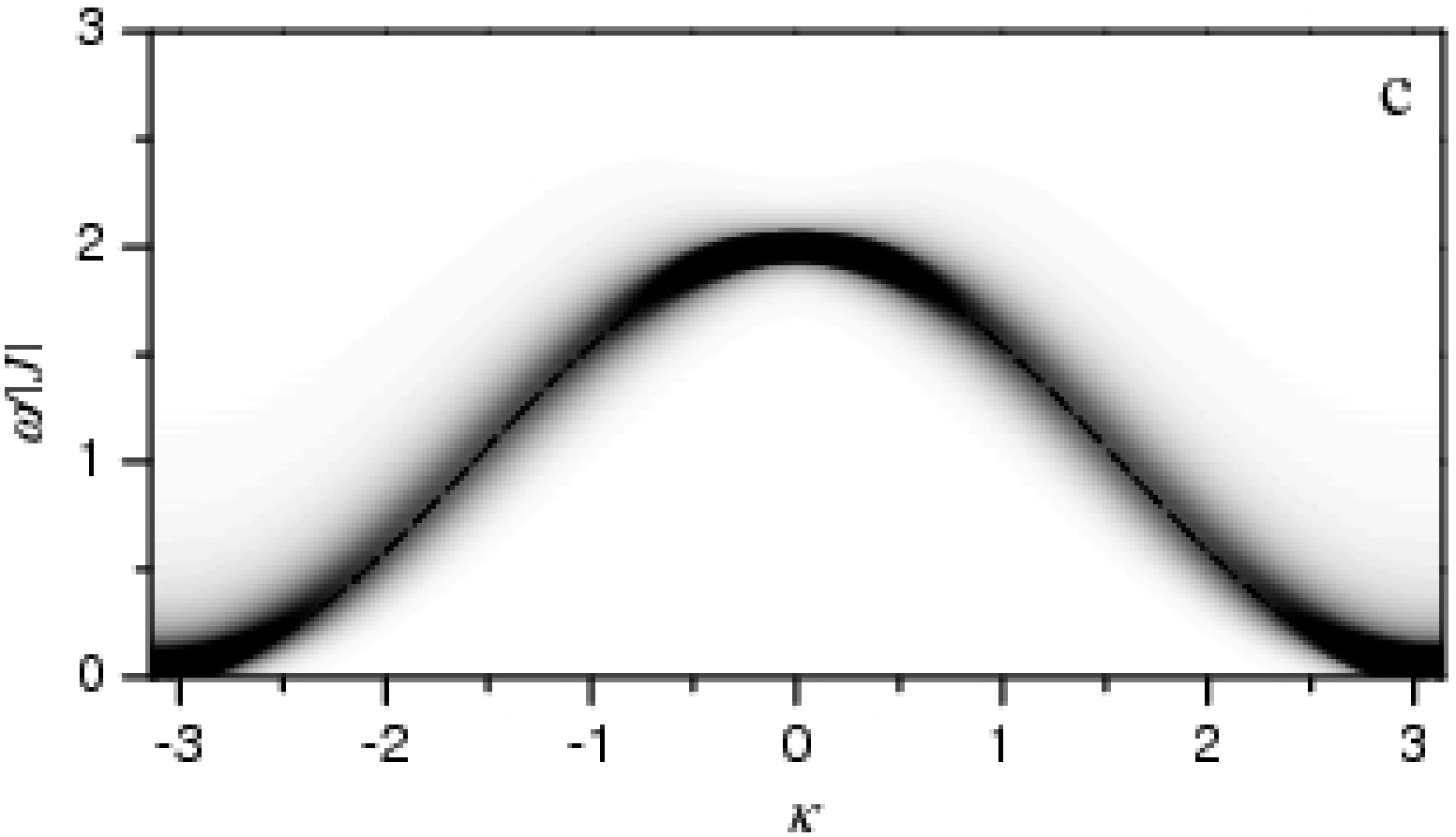}
\includegraphics[clip=on,width=60mm,height=35mm,angle=0]{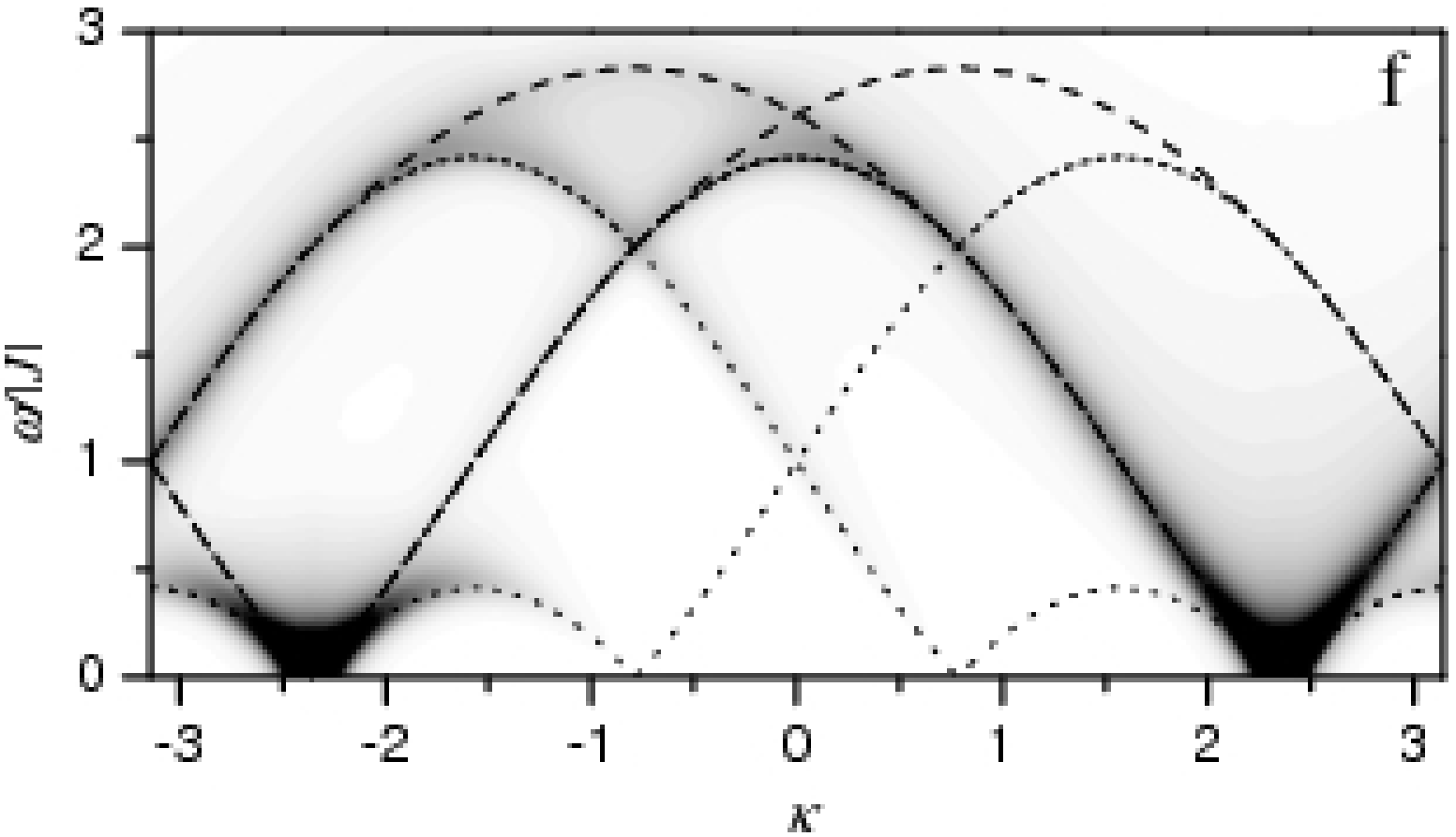}
\end{center}
\caption[]
{\small
$S_{xx}(\kappa,\omega)$
(gray-scale plots)
for the chain (\ref{01}) with $J=1$,
$D=0$ (left panels a, b, c)
and
$D=1$ (right panels d, e, f),
$h=0.001$ (a, d),
$h=0.5$ (b, e),
$h=1$ (c, f)
at the low temperature $\beta=20$.
We have also plotted the boundaries
$\omega^{(1)}(\kappa^{\prime}\pm\varphi)$ (\ref{06}), 
$\omega^{(2)}(\kappa^{\prime}\pm\varphi)$ (\ref{07}) 
(dotted and short-dashed curves)
and
$\omega^{(3)}(\kappa^{\prime}\pm\varphi)$ (\ref{08}) (dashed curves).
\label{fig01}}
\end{figure}

\begin{figure}
\begin{center}
\includegraphics[clip=on,width=60mm,height=35mm,angle=0]{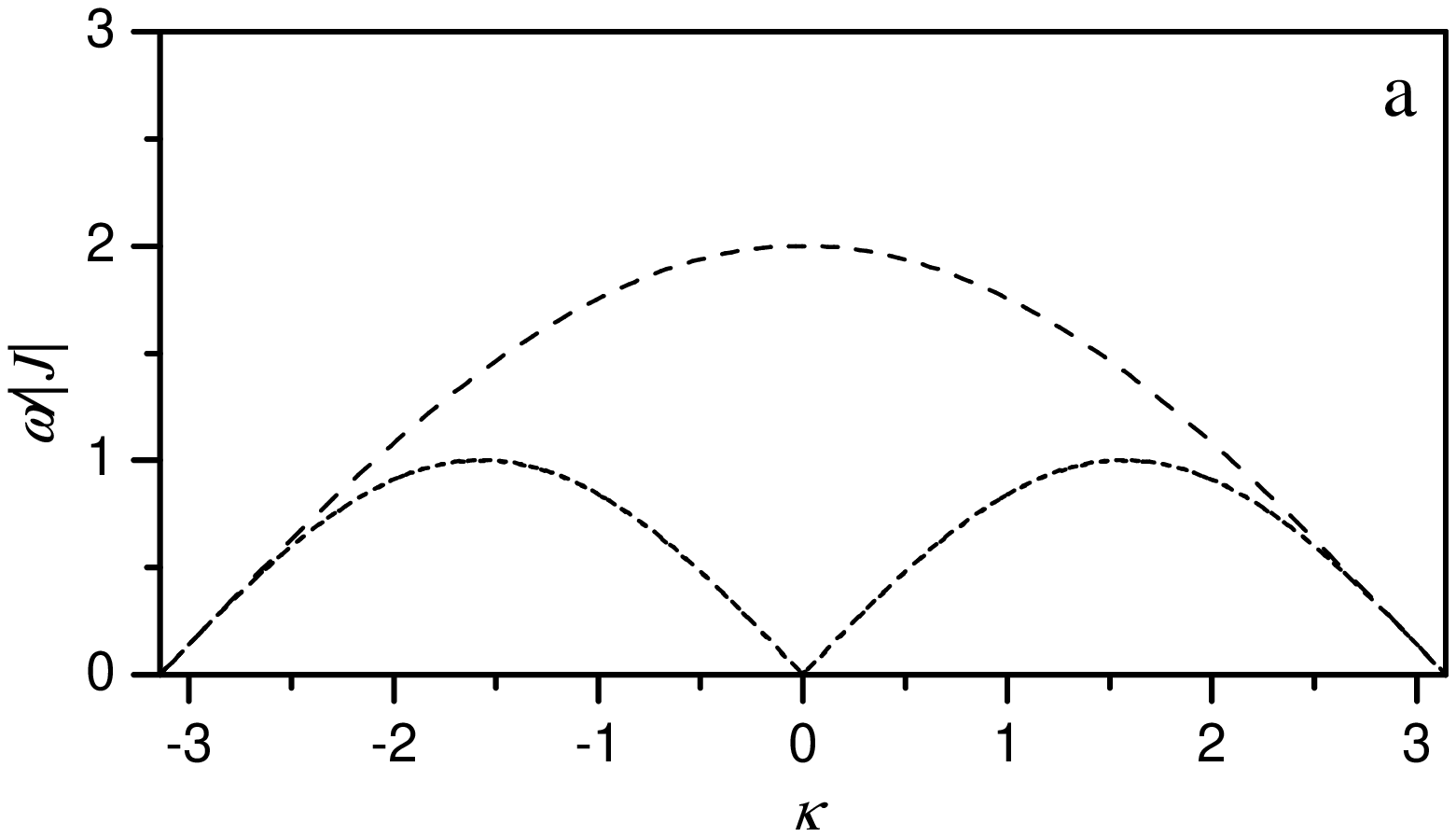}
\includegraphics[clip=on,width=60mm,height=35mm,angle=0]{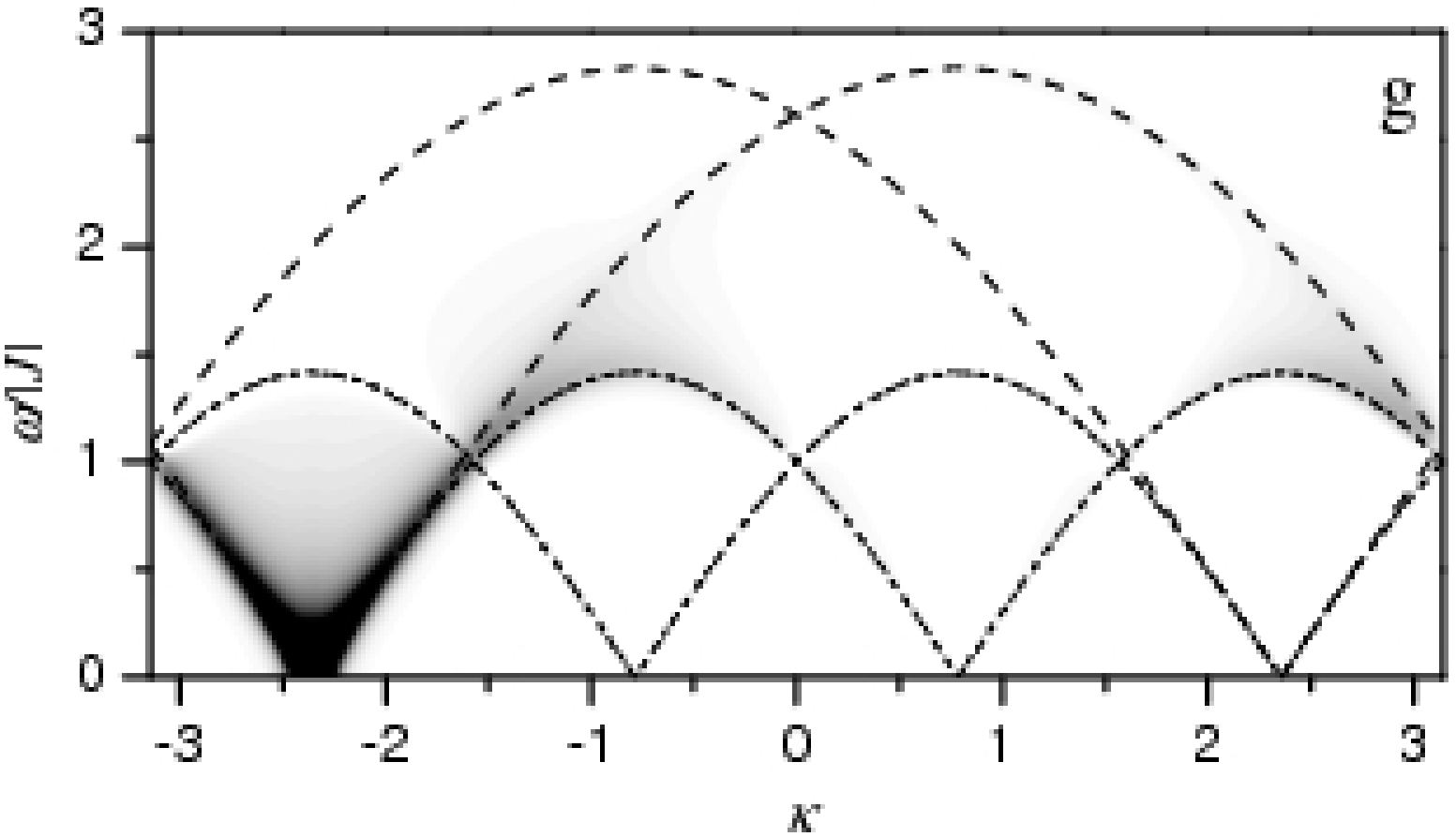}
\linebreak
\includegraphics[clip=on,width=60mm,height=35mm,angle=0]{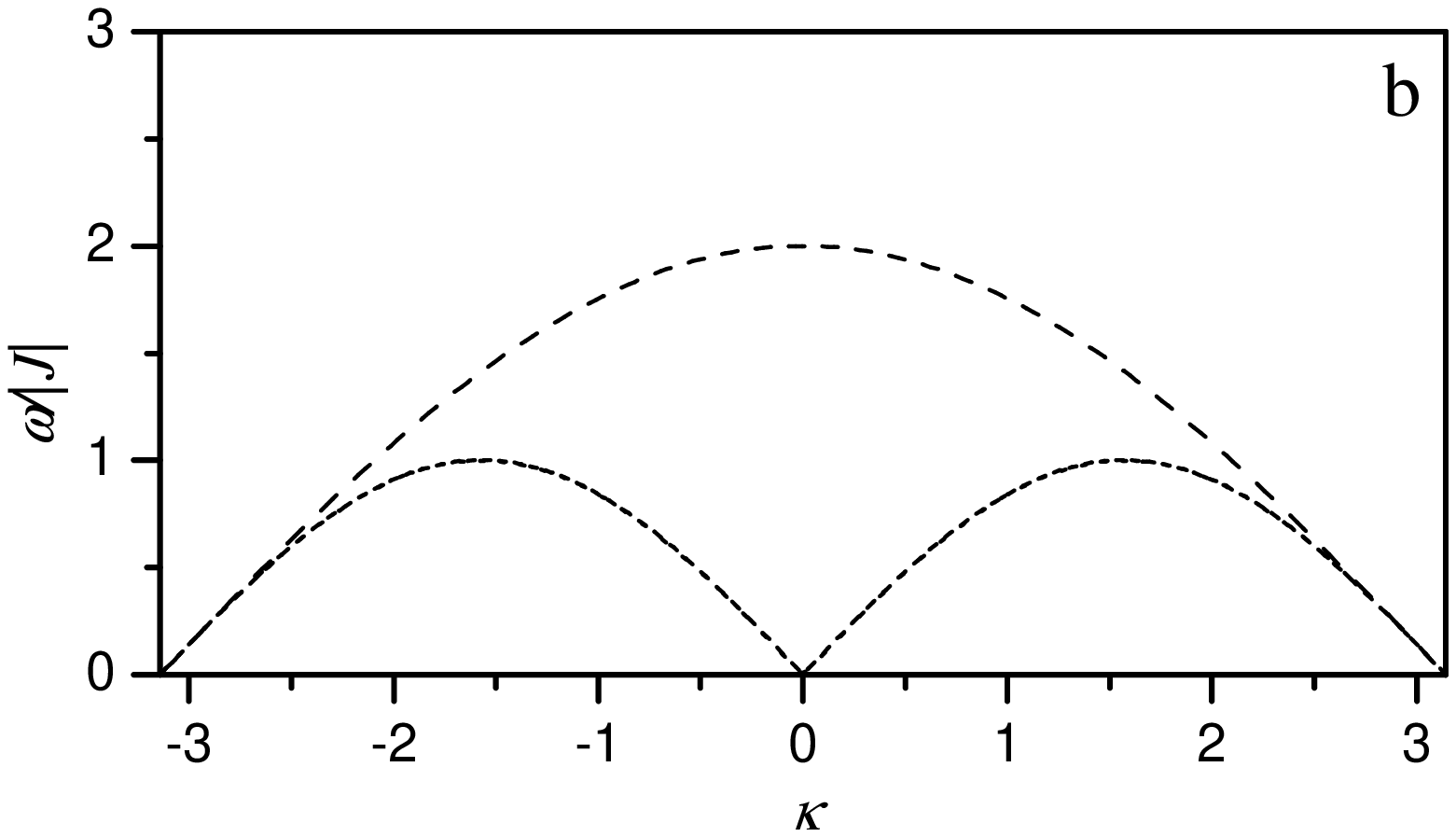}
\includegraphics[clip=on,width=60mm,height=35mm,angle=0]{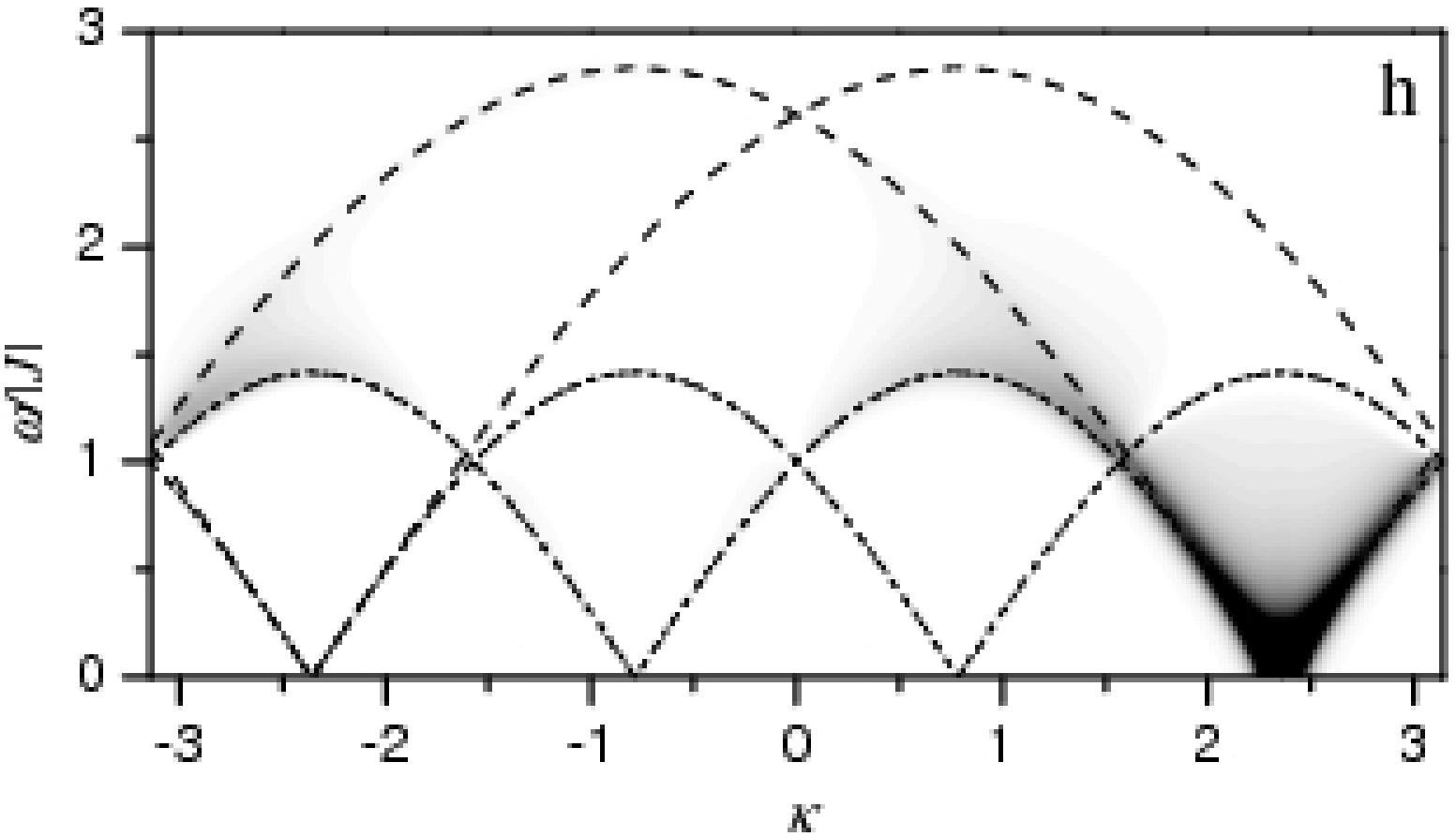}
\linebreak

\vspace{3mm}
\includegraphics[clip=on,width=60mm,height=35mm,angle=0]{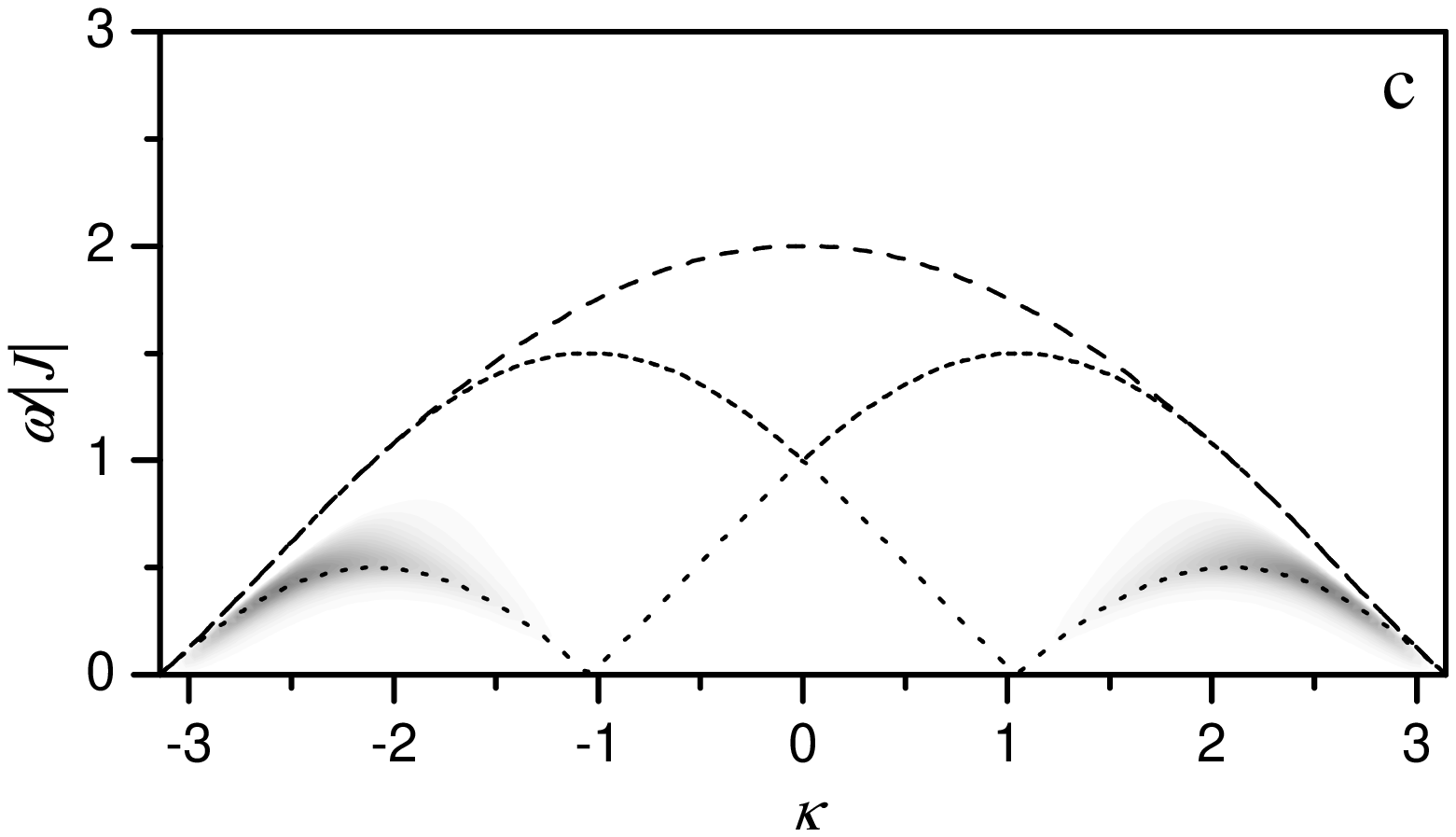}
\includegraphics[clip=on,width=60mm,height=35mm,angle=0]{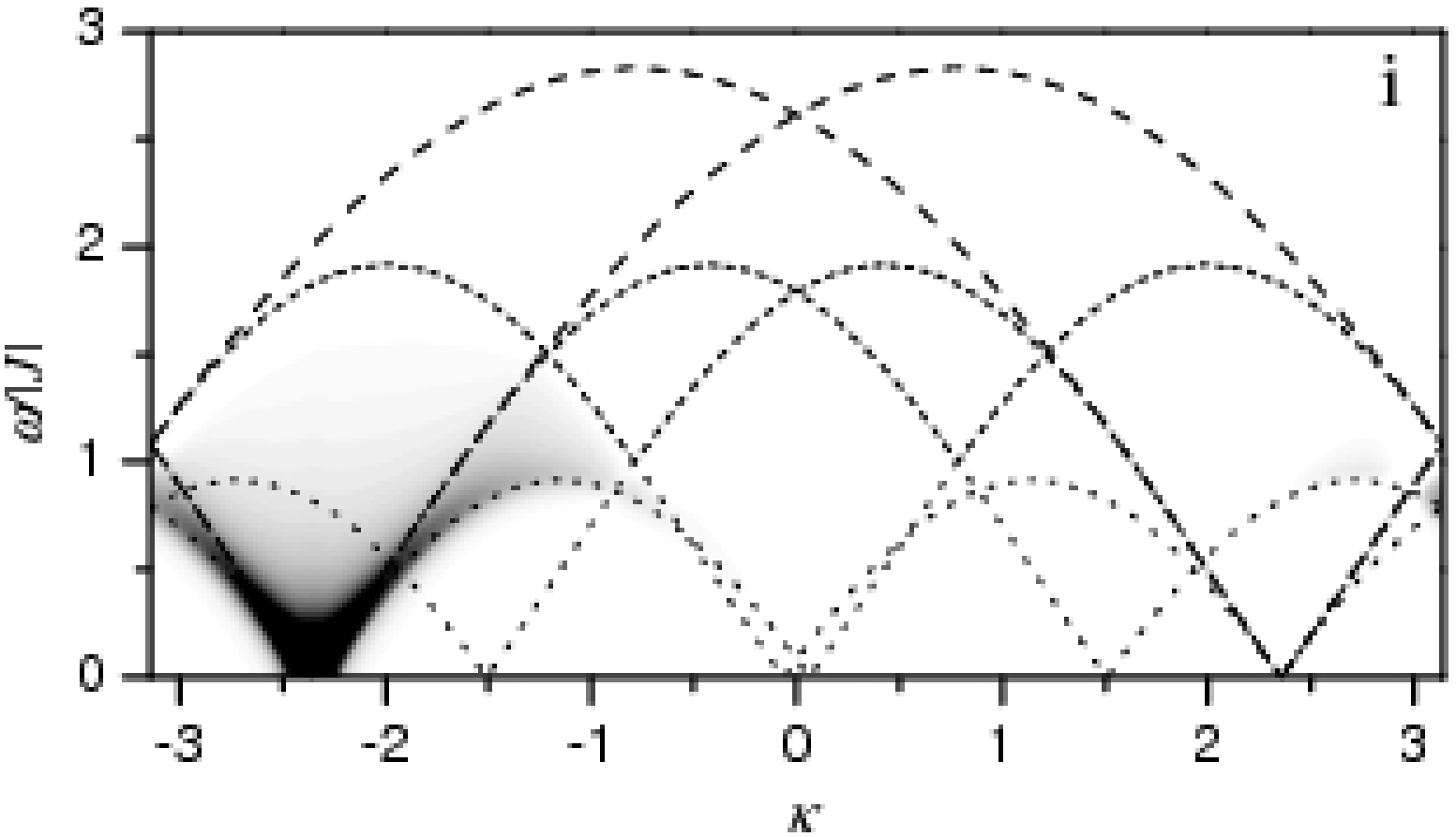}
\linebreak
\includegraphics[clip=on,width=60mm,height=35mm,angle=0]{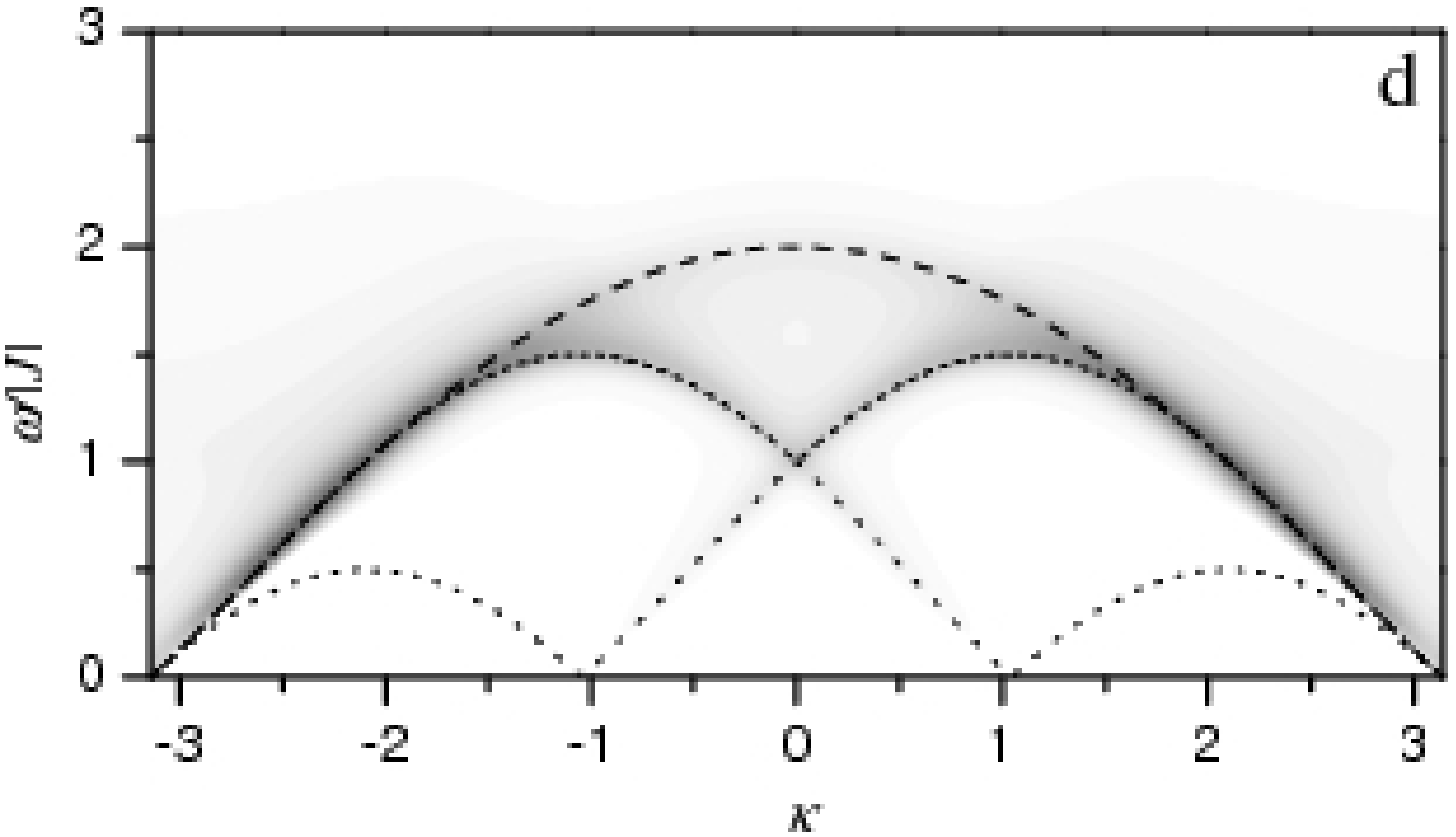}
\includegraphics[clip=on,width=60mm,height=35mm,angle=0]{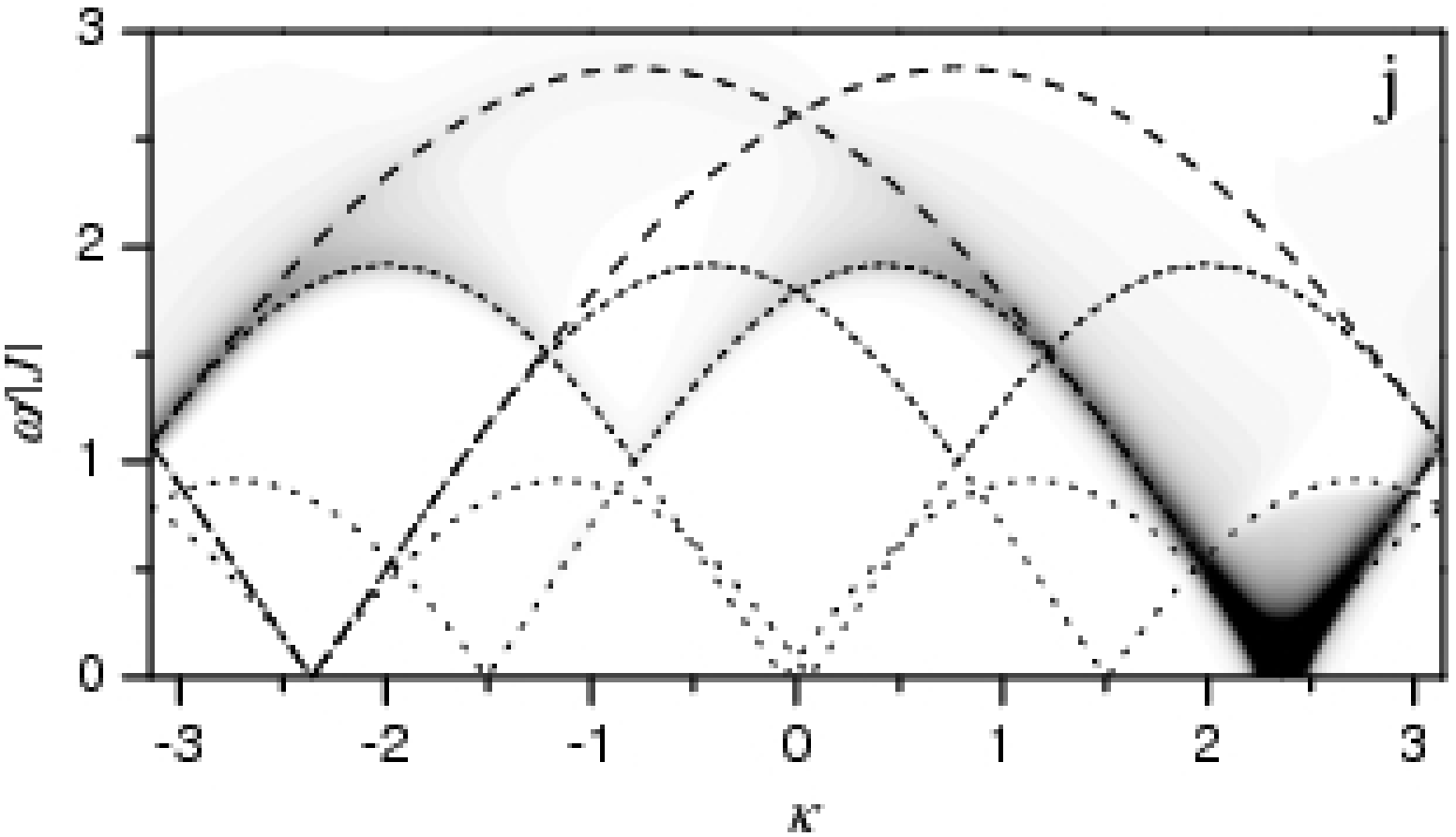}
\linebreak

\vspace{3mm}
\includegraphics[clip=on,width=60mm,height=35mm,angle=0]{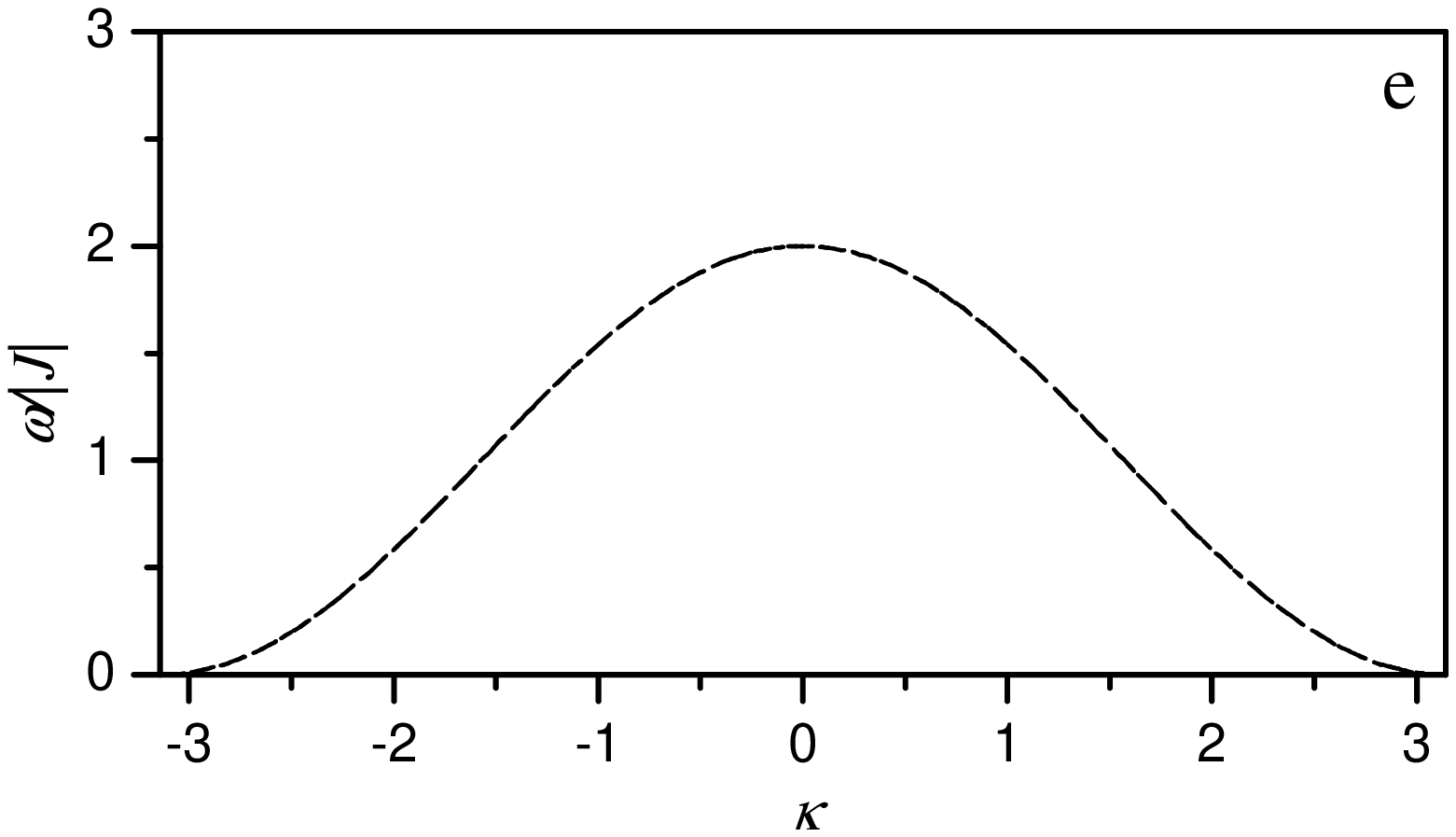}
\includegraphics[clip=on,width=60mm,height=35mm,angle=0]{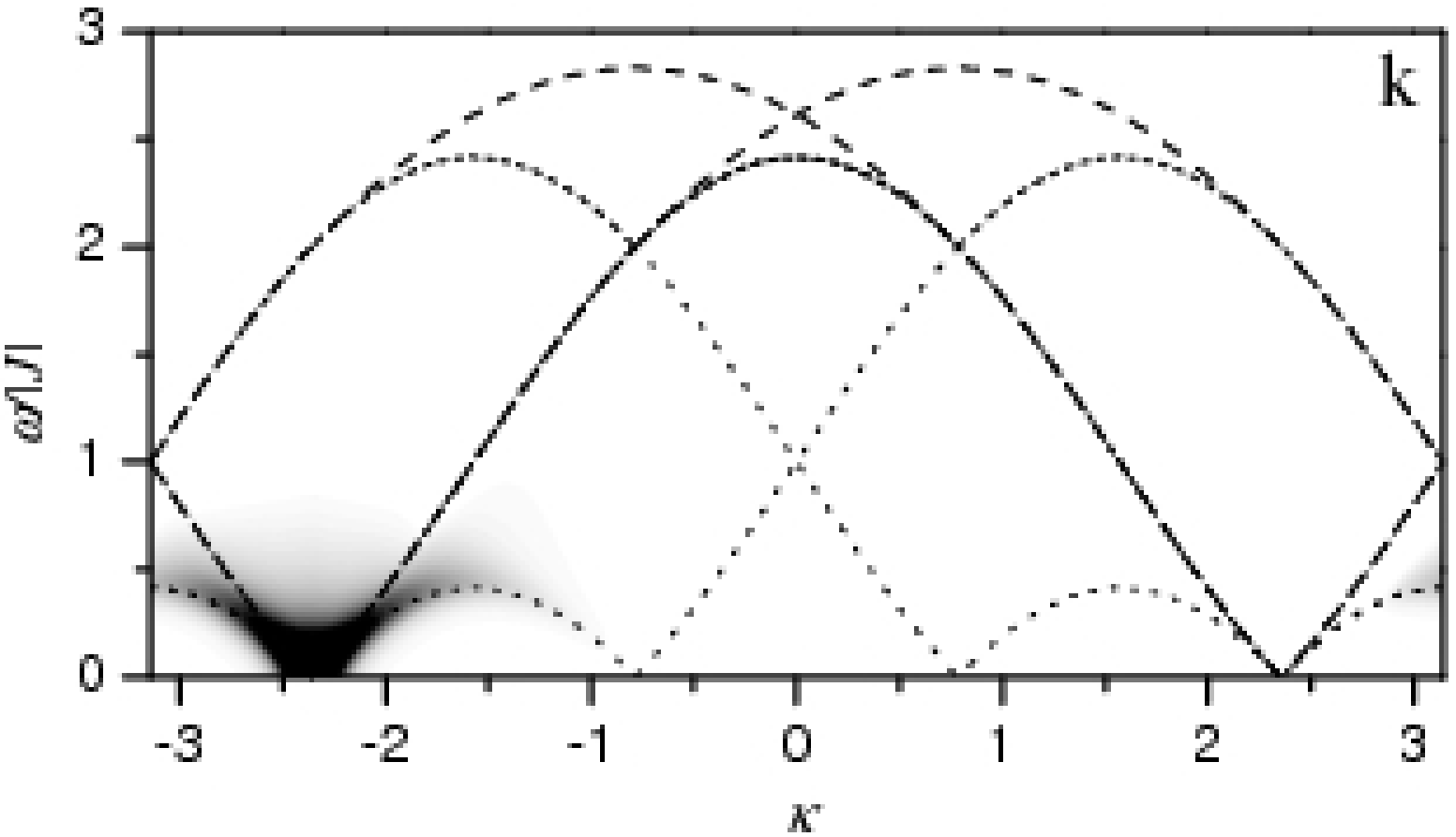}
\linebreak
\includegraphics[clip=on,width=60mm,height=35mm,angle=0]{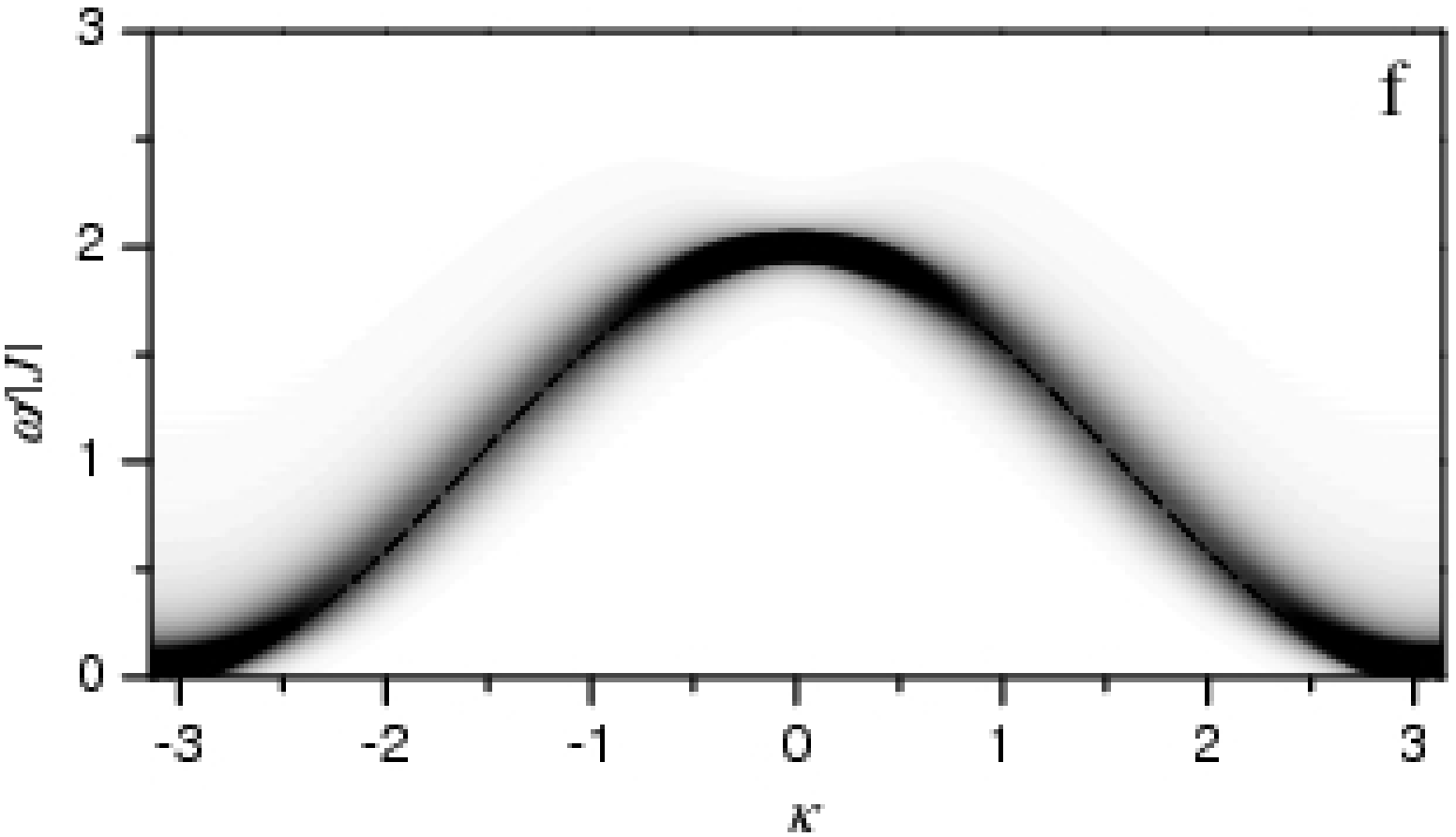}
\includegraphics[clip=on,width=60mm,height=35mm,angle=0]{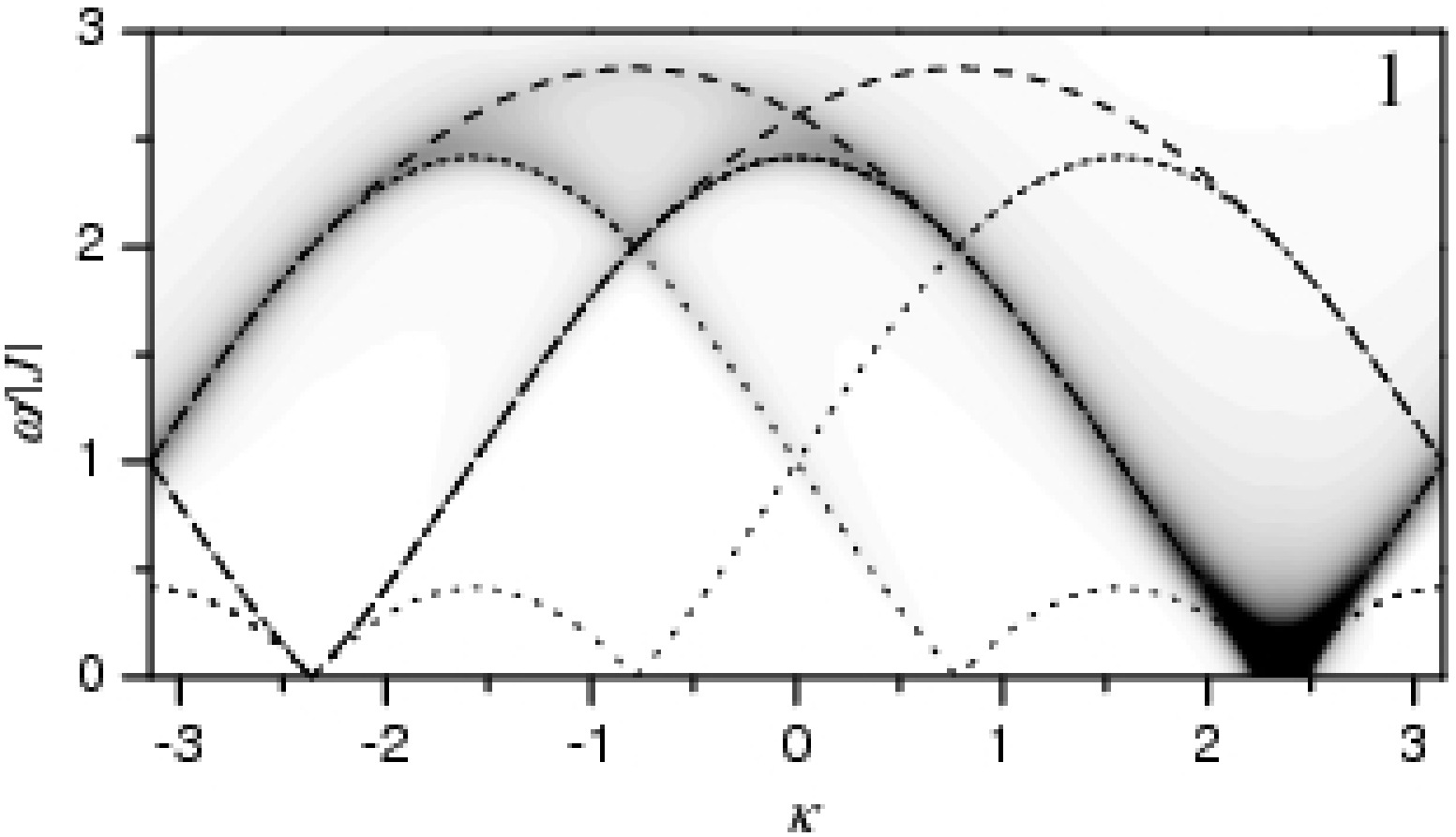}

\end{center}
\caption[]
{\small
The positive (a, g, c, i, e, k)
and negative (b, h, d, j, f, l) parts
of
${\rm{i}}S_{xy}(\kappa,\omega)$
(gray-scale plots)
for the chain (\ref{01}) with $J=1$,
$D=0$ (left panels a, b, c, d, e, f)
and
$D=1$ (right panels g, h, i, j, k, l),
$h=0.001$ (a, b, g, h),
$h=0.5$ (c, d, i, j),
$h=1$ (e, f, k, l)
at the low temperature $\beta=20$.
Note that $S_{xy}(\kappa,\omega)=0$ at $D=0$, $h=0$
(see panels a and b)
and ${\rm{i}}S_{xy}(\kappa,\omega)$ becomes purely negative in the high-field limit
(see panels e, f).
We have also plotted the boundaries
$\omega^{(1)}(\kappa^{\prime}\pm\varphi)$ (\ref{06}),
$\omega^{(2)}(\kappa^{\prime}\pm\varphi)$ (\ref{07}) 
(dotted and short-dashed curves)
and
$\omega^{(3)}(\kappa^{\prime}\pm\varphi)$ (\ref{08}) (dashed curves).
\label{fig02}}
\end{figure}

In the case of finite temperatures $0<\beta<\infty$
we do not have analytical expressions
for the dynamic structure factors 
$S_{xx}(\kappa,\omega)\vert_{\tilde{J}}$ and $S_{xy}(\kappa,\omega)\vert_{\tilde{J}}$
of the model (\ref{04})
but these can be found numerically for chains of several hundreds of sites.
Following the lines explained in detail in Refs. \cite{15a,15c}
we perform the numerical calculations for a chain of $N=400$ sites \cite{26}.
We consider the antiferromagnetic sign of the $XX$ exchange interaction
and fix the units putting $J=1$.
Our results refer to the low temperature $\beta=20$
and they pertain to the thermodynamic limit.
In Figs. \ref{fig01} and \ref{fig02}
we show gray-scale plots
for $S_{xx}(\kappa,\omega)$ and ${\rm{i}}S_{xy}(\kappa,\omega)$,
respectively.
The reported data refer to representative sets of the Hamiltonian parameters
$D=0,\;1$,
$h=0.001,\;0.5,\;1$.
As is nicely seen in Figs. \ref{fig01} and \ref{fig02}
the $xx$ and $xy$ dynamic structure factors are concentrated mostly along
certain lines in the $\kappa$--$\omega$ plane
which are connected with the characteristic lines of the two-fermion excitation continuum
(\ref{06}), (\ref{07}), (\ref{08}).
Let us recall
that in the case $D=0$ and $J<0$
the $xx$ and $xy$ dynamic structure factors
are concentrated roughly along the boundaries of the two-fermion excitation continuum
$\omega_l(\kappa)$, $\omega_m(\kappa)$, $\omega_u(\kappa)$
\cite{15c}.
In the case $D=0$ and $J>0$
(see the left panels in Figs. \ref{fig01} and \ref{fig02})
the symmetry relation mentioned in Sec. \ref{s2} implies
that these dynamic structure factors
should be concentrated roughly along the lines
$\omega_l(\kappa^{\prime})=\omega_l(\kappa\pm\pi)$,
$\omega_m(\kappa^{\prime})=\omega_m(\kappa\pm\pi)$,
$\omega_u(\kappa^{\prime})=\omega_u(\kappa\pm\pi)$.
That is really the case
as can be seen in the left panels in Figs. \ref{fig01} and \ref{fig02}.
In the presence of the Dzyaloshinskii-Moriya interaction $D\ne 0$
this simple correspondence is violated
becoming more intricate.
Namely,
the two-fermion excitation continuum splits into two continua,
the ``left'' one
(with the boundaries
$\omega_l(\kappa^{\prime}-\varphi)$,
$\omega_m(\kappa^{\prime}-\varphi)$,
$\omega_u(\kappa^{\prime}-\varphi)$)
and
the ``right'' one
(with the boundaries
$\omega_l(\kappa^{\prime}+\varphi)$,
$\omega_m(\kappa^{\prime}+\varphi)$,
$\omega_u(\kappa^{\prime}+\varphi)$);
these continua are connected by the symmetry operations
discussed in Sec. \ref{s2}.
The larger $D$ is 
the larger is the splitting
controlled by the value of $\varphi$.
At fixed $D\ne 0$ and $h=0$
the spectral weight is equally distributed between the left and the right continua,
resulting in a symmetry with respect to $\kappa\to -\kappa$
(Figs. \ref{fig01}d and \ref{fig02}g, \ref{fig02}h).
Asymmetry with respect to $\kappa\to -\kappa$
arises as $h$ deviates from zero.
While $\vert h\vert$ increases from 0 to $\sqrt{J^2+D^2}$
the spectral weight ``moves'' from the left continuum to the right one
and all the spectral weight becomes concentrated along the boundaries of the right continuum
as $\vert h\vert$ approaches $\sqrt{J^2+D^2}$
(Figs. \ref{fig01}f and \ref{fig02}k, \ref{fig02}l).
The presence of the Dzyaloshinskii-Moriya interaction
produces a number of specific changes
nicely seen in the right panels in Figs. \ref{fig01}, \ref{fig02}.
For example,
the field-independent modes at $\pm\pi$ shift to $\pm\pi\mp\varphi$.
One can also see noticeable changes
in the frequency profiles for fixed value of the wavevector
(say,
$\kappa=0$ or $\kappa=\pm\pi$).

Finally,
we turn to the case of zero temperature $\beta\to\infty$.
In this case the exact result for strong fields,
$\vert h\vert>\sqrt{J^2+D^2}$ is known \cite{12}.
Since the ground state is completely polarized
(i.e. completely empty/filled in fermionic language)
the time-dependent spin correlation functions
$\langle s_n^x(t) s_{n+m}^x\rangle\vert_{\tilde{J}}$
and
$\langle s_n^x(t) s_{n+m}^y\rangle\vert_{\tilde{J}}$
can be easily calculated \cite{12}.
That yields the following results for the dynamic structure factors (\ref{02})
of the model (\ref{01})
in this limit
\begin{eqnarray}
S_{xx}(\kappa,\omega)
=-{\rm{sgn}}(h)\; {\rm{i}}S_{xy}(\kappa,\omega)
=\frac{\pi}{2}
\delta\left(\omega-\vert h\vert-\tilde{J}\cos(\kappa+{\rm{sgn}}(h)\varphi)\right).
\label{13}
\end{eqnarray}
In accordance with Eq. (\ref{13})
${\rm{sgn}}\left({\rm{i}}S_{xy}(\kappa,\omega)\right)
=-{\rm{sgn}}\left(h\right)$
and ${\rm{i}}S_{xy}(\kappa,\omega)<0$ for positive $h$
(actually for $h>\sqrt{J^2+D^2}$)
(see the low-temperature data in Fig. \ref{fig02}, panels e, f).
Further,
all of the spectral weight for both dynamic quantities
is concentrated along the curve
\begin{eqnarray}
\frac{\omega^{\star}(\kappa)}{\sqrt{J^2+D^2}}
=
\frac{\vert h\vert}{\sqrt{J^2+D^2}}
+
{\rm{sgn}}(J)\cos\left(\kappa+{\rm{sgn}}(h)\varphi\right).
\label{14}
\end{eqnarray}
As $\vert h\vert\to \sqrt{J^2+D^2}$
the r.h.s. of Eq. (\ref{14}) transforms
either into $2\cos^2((\kappa+{\rm{sgn}}(h)\varphi)/2)$
if $J>0$
(compare with the low-temperature data
reported in Figs. \ref{fig01}c and \ref{fig02}f)
or into $2\sin^2((\kappa+{\rm{sgn}}(h)\varphi)/2)$
if $J<0$.

Having calculated
(partly analytically and partly numerically)
the $xx$ and $xy$ dynamic structure factors of the model (\ref{01}),
we conclude
that these dynamic quantities
away from the infinite temperature limit $\beta=0$
exhibit a number of peculiar features,
namely,
asymmetry with respect to $\kappa\to -\kappa$ at $h\ne 0$,
specific structure of frequency profiles at fixed values of $\kappa$,
field-independent positions of soft modes,
which can be used for unambiguous determination of the Dzyaloshinskii-Moriya interaction.

\section{ESR absorption spectrum in the presence of Dzyaloshinskii-Moriya interaction}
\label{s5}

Our results for the dynamic structure factors may be used to discuss
the effect of the Dzyaloshinskii-Moriya interaction
on the energy absorption
in electron spin resonance (ESR) experiments.
We notice,
a similar analysis of ESR in the spin-$1/2$ $XX$ chain
is reported in \cite{04}.
Clearly,
that kind of analysis can be extended to the case
when the Dzyaloshinskii-Moriya interaction is present.

\begin{figure}
\begin{center}
\includegraphics[clip=on,width=65mm]{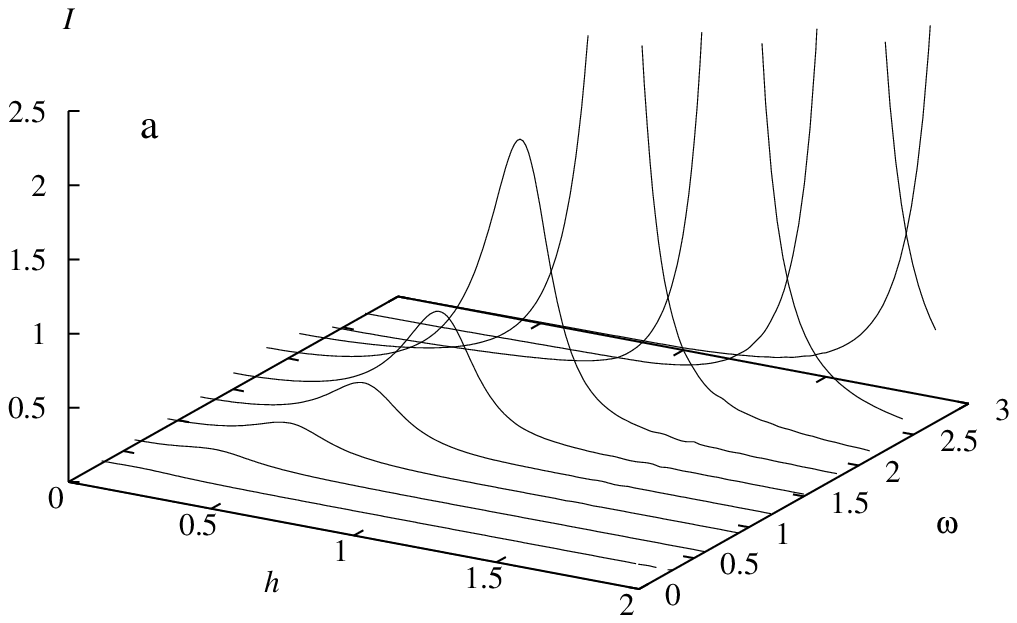}
\includegraphics[clip=on,width=65mm]{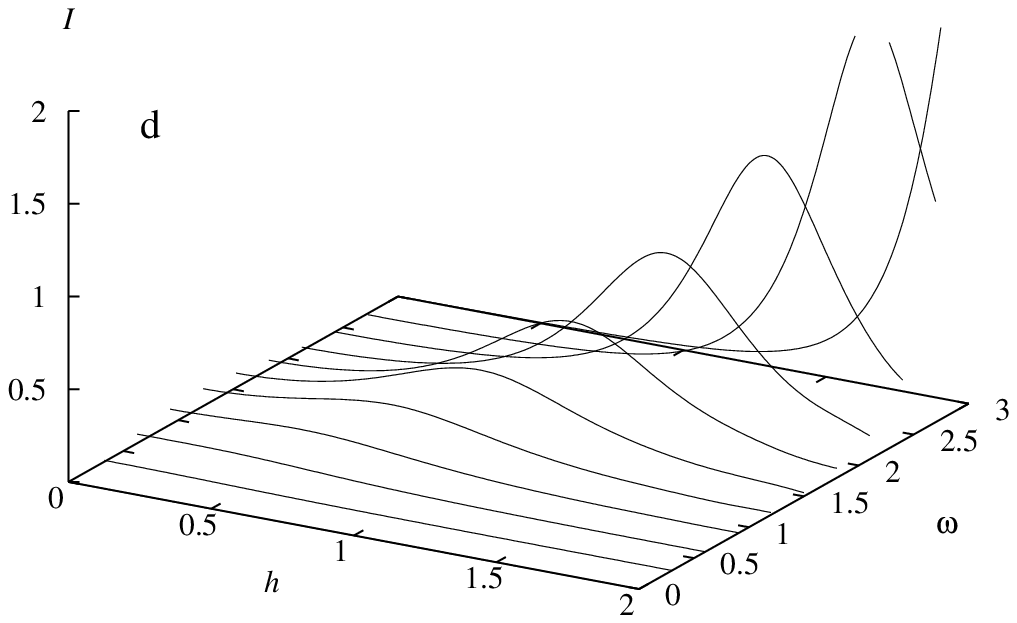}
\linebreak

\vspace{4mm}
\includegraphics[clip=on,width=65mm]{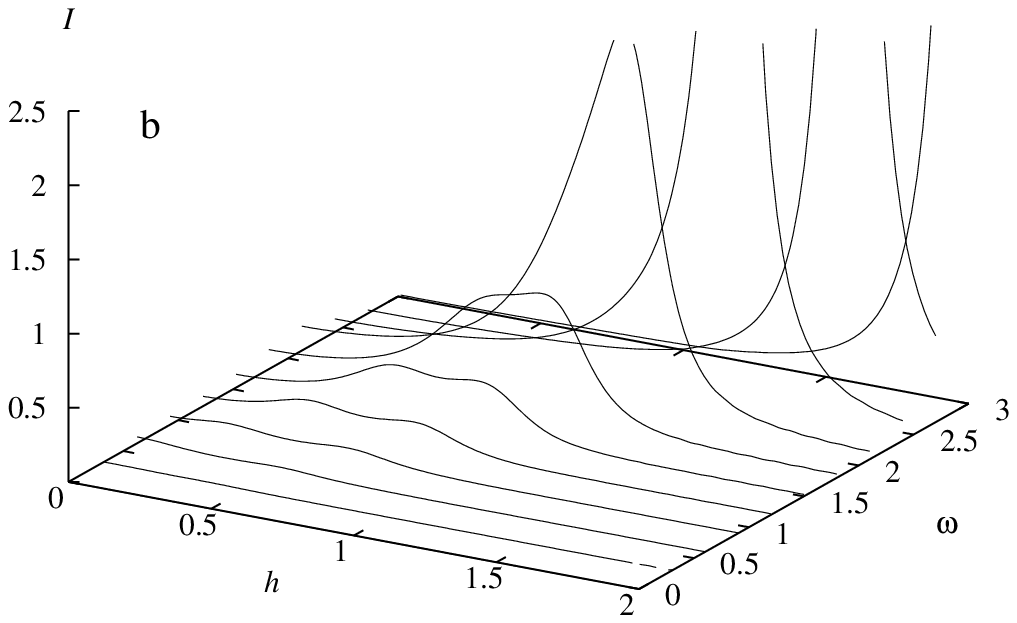}
\includegraphics[clip=on,width=65mm]{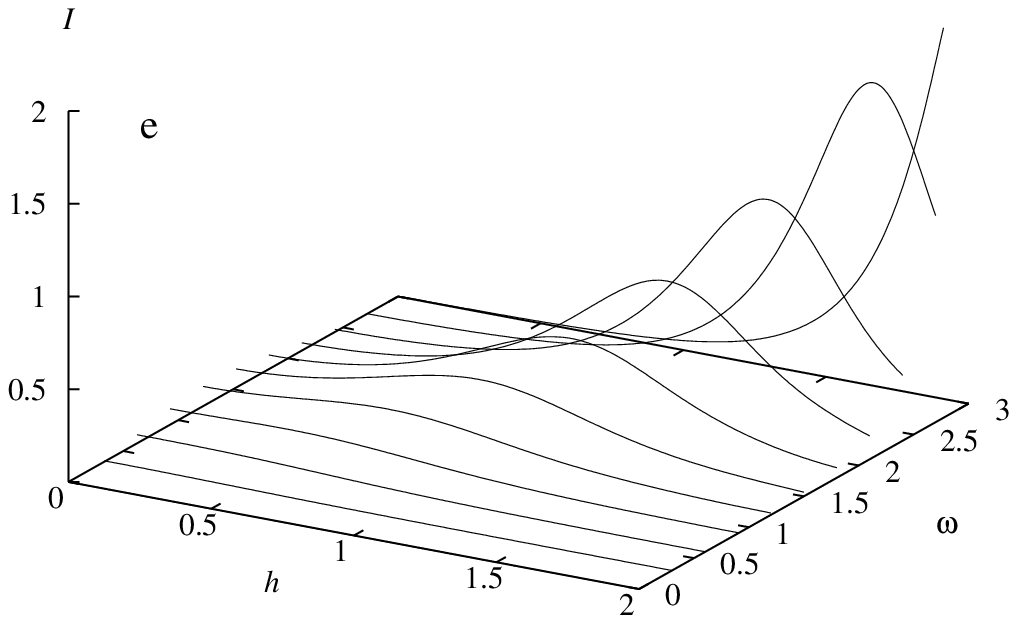}
\linebreak

\vspace{4mm}
\includegraphics[clip=on,width=65mm]{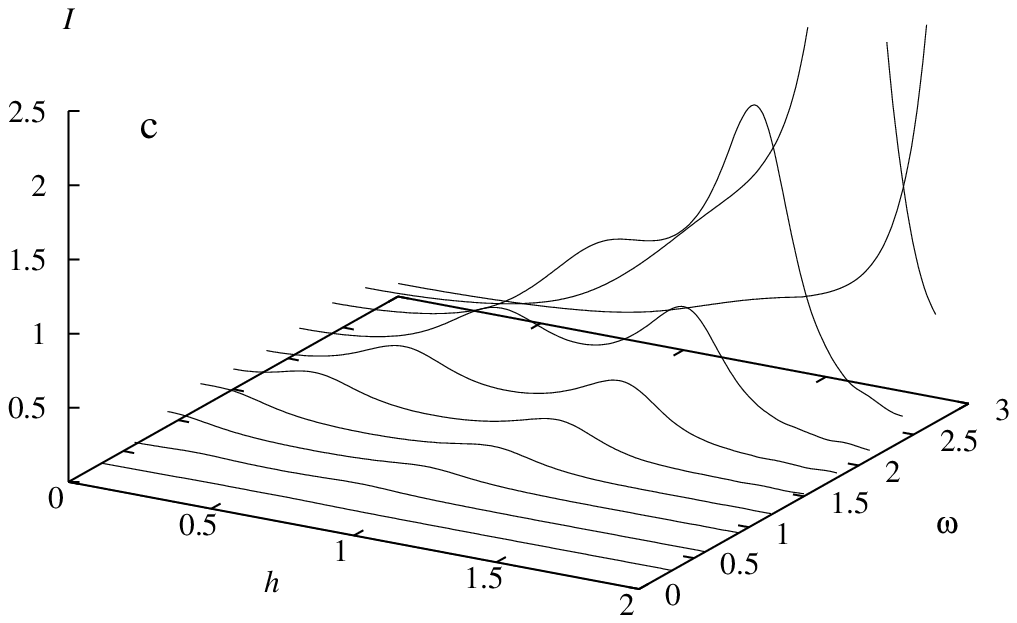}
\includegraphics[clip=on,width=65mm]{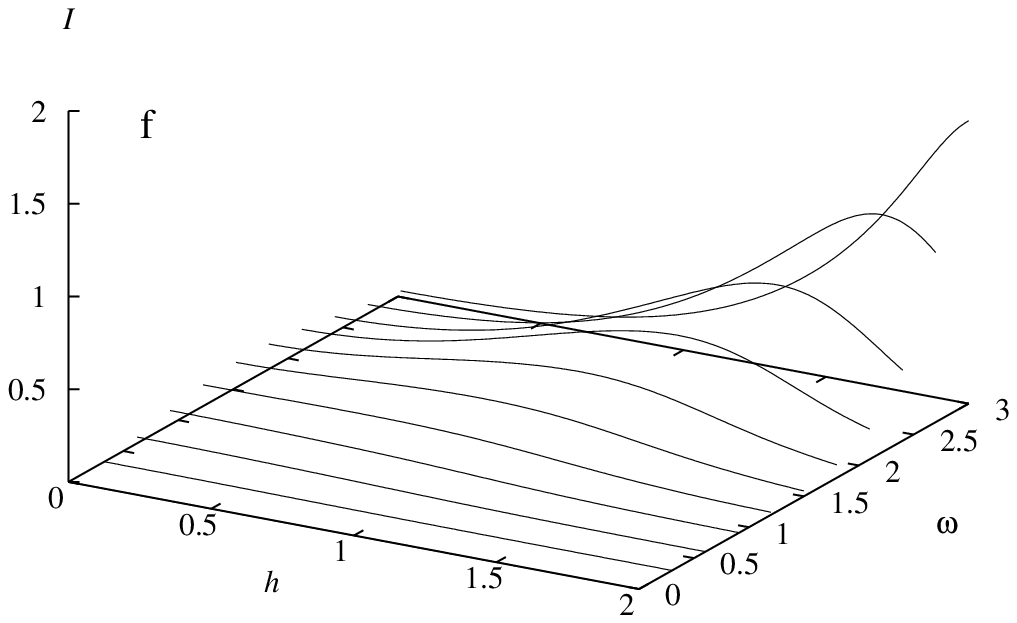}
\end{center}
\caption[]
{\small
The low-temperature (left column a, b, c)
and
the intermediate-temperature (right column d, e, f)
absorption intensity $I$ (\ref{15}), (\ref{17})
at different frequencies $\omega$ and magnetic fields $h$
for the spin-$1/2$ $XX$ chain with the Dzyaloshinskii-Moriya interaction.
$J=1$,
$D=0$ (a, d),
$D=0.5$ (b, e),
$D=1$ (c, f),
$\beta=5$ (left panels a, b, c),
$\beta=1$ (right panels d, e, f).
\label{fig03}}
\end{figure}

Consider an ESR experiment,
in which
the static magnetic field directed along the $z$ axis
and
the electromagnetic wave with the polarization in the $\alpha\perp z$ direction
(say, $\alpha=x$)
are applied to a magnetic system
which is described as a spin-$1/2$ $XX$ chain with Dzyaloshinskii-Moriya interaction
(ESR experiment in the standard Faraday configuration).
In such an ESR experiment
one measures
the intensity of the radiation absorption $I(\omega)$
as a function of frequency $\omega>0$
of the electromagnetic wave.
Within the linear response theory
the absorption intensity is written as
\begin{eqnarray}
I(\omega)
\propto
\omega\Im\chi_{\alpha\alpha}(0,\omega),
\label{15}
\end{eqnarray}
where
$\Im\chi_{\alpha\alpha}(0,\omega)$
is the imaginary part
of the ($\alpha\alpha$) diagonal component of the dynamic susceptibility
\begin{eqnarray}
\chi_{\alpha\beta}(\kappa,\omega)
=
{\rm{i}}\sum_{m=1}^N\exp\left(-{\rm{i}}\kappa m\right)
\int_0^{\infty}{\rm{d}}t\exp\left({\rm{i}}\left(\omega+{\rm{i}}\epsilon\right)t\right)
\langle\left[ s_{n+m}^{\alpha}(t), s_n^{\beta}\right]\rangle,
\;\;\;
\epsilon\to +0
\label{16}
\end{eqnarray}
at zero wavevector $\kappa=0$
(see e.g. \cite{03}).
We notice that
\begin{eqnarray}
\Im\chi_{\alpha\alpha}(0,\omega)
=
\frac{1-\exp(-\beta\omega)}{2}S_{\alpha\alpha}(0,\omega),
\label{17}
\end{eqnarray}
where $S_{\alpha\alpha}(0,\omega)$ is given by (\ref{02})
(see \cite{04,27})
and hence
our findings for $S_{xx}(\kappa,\omega)$
are directly related to the ESR absorption $I(\omega)$
for the spin-$1/2$ $XX$ chain with the Dzyaloshinskii-Moriya interaction.
Further we restrict ourselves to the antiferromagnetic sign of the $XX$ exchange interaction
$J>0$.
In Fig. \ref{fig03}
we report a typical dependence
of the absorption intensity $I$ (\ref{15}), (\ref{17}) at fixed frequency $\omega$
on the applied static magnetic field $h$
(three-dimensional plots)
obtained numerically
for a chain of $N=400$ sites \cite{28}.
The results refer to low (left) and intermediate (right) temperatures.
They demonstrate the changes in the absorption intensity as $D$ increases (from top to bottom).

We now discuss the effect of the Dzyaloshinskii-Moriya interaction
on the ESR absorption intensity
for the spin-$1/2$ $XX$ antiferromagnetic chain.
We start from {\bf{the strong-field limit}} $h>\sqrt{J^2+D^2}$
when all interspin interactions can be regarded as perturbations.
At zero temperature
according to (\ref{15}), (\ref{17}), (\ref{13})
we have
$I(\omega)\propto(\pi/4)\omega\delta\left(\omega-h-J\right)$.
Thus,
for $h>\sqrt{J^2+D^2}$ and $\beta\to\infty$
the resonance is completely sharp and located exactly at $h+J$.
Remarkably,
the Dzyaloshinskii-Moriya interaction drops out of the absorption intensity!
The behavior of the absorption intensity in the strong-field limit
for low and intermediate temperatures
can be read off the left and right panels in Fig. \ref{fig03}, respectively.
For the infinite-temperature limit $\beta=0$
we have
(for arbitrary fields)
$I(\omega)\propto (1/2)\left(\beta\omega^2+{\cal{O}}(\beta^2)\right)S_{xx}(0,\omega)$
with
$S_{xx}(0,\omega)$ given by Eq. (\ref{11})
and hence $I$ displays a Gaussian peak at $\omega=h$.

The simple picture valid in the strong-field limit breaks down
{\bf{below the saturation field}}
$0<h<\sqrt{J^2+D^2}$.
In this case the interspin interactions become important
and we find more complicated behaviors
(see Fig. \ref{fig03}).
In particular,
in the presence of the Dzyaloshinskii-Moriya interaction
the $h$-dependence of the absorption intensity at low temperatures
in a certain frequency range
may show a two-peak structure
(compare Figs. \ref{fig03}a, \ref{fig03}b, \ref{fig03}c).
This is clearly connected
to the above discussed splitting of the two-fermion excitation continuum,
along the characteristic lines of which
the $xx$ dynamic structure factor is mostly concentrated.

Obviously,
one may perform an exhaustive analysis of the ESR absorption
examining the detailed structure of the frequency/field profiles,
the resonant shift, the linewidth etc.
A complete theoretical description of ESR
in the spin-$1/2$ $XX$ chain with the Dzyaloshinskii-Moriya interaction
is left for further studies.
Unfortunately,
we are not aware of ESR experiments on spin-$1/2$ $XX$ chain materials.
However,
there are now several suitable magnetic materials
(see e.g. \cite{29,30} and also \cite{31,32,33})
in which the presence of the Dzyaloshinskii-Moriya interaction cannot
{\it{a priori}}
be excluded
and hence we may expect
that some features described above
may be observed in future ESR experiments.
We hope the present work
to stimulate further theoretical and experimental studies on this subject.

\section{Concluding remarks}
\label{s6}

To summarize,
we have presented a comprehensive treatment of all dynamic structure factors
of the spin-$1/2$ $XX$ chain in a transverse field
with the Dzyaloshinskii-Moriya interaction
providing explicit analytical expressions and high-precision numerical data.
We have shown that the $xx$ ($yy$), $xy$ ($yx$) dynamic quantities
may be used for determining the value of the Dzyaloshinskii-Moriya interaction.
We have discussed briefly
the ESR absorption spectrum
for the spin-$1/2$ $XX$ chain with the Dzyaloshinskii-Moriya interaction.

In recent years,
following the progress in material sciences
the interest in quantum spin chain compounds has noticeably increased.
We notice that some of those compounds
are good realizations of the one-dimensional spin-$1/2$ $XX$ model
(e.g. Cs$_2$CoCl$_4$ has been proposed as a possible quasi-one-dimensional $XY$-like magnet)
\cite{29,30,31,32,33}.
Dynamic experiments are an important tool to investigate these compounds.
Neutron scattering, ESR, NMR (see e.g. the recent paper \cite{34}) etc.
yield experimental probes of the dynamic properties
which have to be compared with theoretical predictions.
In the present work
within the frame of the simple model
we follow rigorously 
how the Dzyaloshinskii-Moriya interaction manifests itself
in the quantities accessible to experimental investigation.

As a final remark,
we emphasize that the dynamics
of the spin-$1/2$ anisotropic $XY$ chain
with the Dzyaloshinskii-Moriya interaction is much more involved \cite{35}
since the Dzyaloshinskii-Moriya interaction
cannot be eliminated by a spin axes transformation.
The detailed study of the effects of the Dzyaloshinskii-Moriya interaction
on the dynamic quantities
in the presence of anisotropy of the $XY$ exchange interaction
is in progress.

\section*{Acknowledgments}
\label{s7}
The authors thank 
Taras Krokhmalskii
for collaboration on closely related topics and numerical assistance,
Joachim Stolze
for helpful comments on the manuscript
and 
Brian Rainford
for useful correspondence.
O.~D. is grateful to the ICTP, Trieste
for the kind hospitality
in the autumn of 2005
when the paper was completed.
The paper was partially presented
at the International Conference on Strongly Correlated Electron Systems SCES '04
(July 26 - 30, 2004,
Karlsruhe, Germany).
O.~D. thanks the organizers
for the grant for participation in the conference.


\begin{thebibliography}{100}
\bibitem{26}
In our numerical computations
we take $n=51$ in Eq. (\ref{02})
and make the summation over 100 neighbors.
We also multiply the integrand by $\exp(-\varepsilon\vert t\vert)$
with $\varepsilon=0.01$
to remove the effect of the finite upper limit of integration in Eq. (\ref{02}),
$t_c=300$
(see \cite{15a,15c} for more details).
\bibitem{28}
In our numerical computations
we take $n=51,\;81$ in Eq. (\ref{02}),
make the summation over up to 100 neighbors 
and slightly increase $\varepsilon$ (up to 0.15 for $D=1$) (see \cite{26})
to suppress the time oscillations
which are still present at $t_c=150 \ldots 300$
if $h>\sqrt{J^2+D^2}$.
\end{thebibliography}
\end{document}